\documentclass[reprint,prb,superscriptaddress,amsmath,amssymb]{revtex4-2}

\usepackage{graphicx}
\usepackage{bm}
\usepackage[colorlinks=true,allcolors=blue]{hyperref}
\usepackage{xcolor}

\newcommand{\FigSuppBoundaries}{S3}

\newcommand{\FigSuppHoneCycle}{S1}
\newcommand{\FigSuppHoneTau}{S2}
\newcommand{\TabSuppWalls}{S3}

\newif\ifarxiv
\arxivtrue   % \arxivfalse para la versión de PRB

\newcommand{\rev}[1]{{#1}}

\begin{document}

\title{\rev{Zeroth persistent} homology of ground-state correlations at clean and disordered quantum transitions}

\author{Francisco Belch\'{\i}}
%\email{francisco.belchi@salle.url.edu}
\affiliation{Universitat Ramon Llull, La Salle Campus Barcelona, Human-Environment Research (HER), Spain}

\author{Nahuel Statuto}
\email{nstatuto@ub.edu}
\affiliation{Departament de Matemàtiques i Informàtica, Universitat de Barcelona, Barcelona, Spain}

\begin{abstract}
The antiferromagnetic Ising chain in transverse and longitudinal fields exhibits a continuous critical point, a mixed-order transition line and, with quenched disorder, an infinite-disorder fixed point and reentrant order. Using exact free-fermion solutions and the density-matrix renormalization group, we examine what the zeroth persistent homology of staggered ground-state correlations adds to conventional responses. Its barcode records the merge scales of the single-linkage hierarchy, and the maximum of its persistence-landscape velocity defines a threshold-free pseudocritical point. This point locates the clean critical point and the mixed-order line with an accuracy comparable to conventional responses. In the reentrant regime, the global velocity misplaces the entry, which is recovered only by restricting the filtration to the ordering scale, and none of the tested estimators locates the exit. A domain wall forced by the boundary conditions suppresses the disorder-averaged reentrant lobe. For individual disordered realizations, the labelled merge tree recovers most strong-disorder renormalization-group clusters, including noncontiguous ones missed by lattice-restricted hierarchies, with increasing agreement at stronger disorder. This information resides in the full correlation matrix, while zeroth homology provides a stable, threshold-free hierarchy.
\end{abstract}

\maketitle

\section{Introduction}
\label{sec:intro}

A quantum phase transition reorganizes the correlations of the ground state.
In a finite system it appears as a rounded feature of some response, and its
location defines a pseudocritical point that approaches the critical point as
the size grows~\cite{FisherBarber1972,Barber1983}. Which response to follow
depends on the transition. The order parameter and its susceptibility work
when the order is known, disordered systems require the statistics of
pseudocritical points over realizations~\cite{AharonyHarris1996,
WisemanDomany1998,IgloiLinRiegerMonthus2007}, and transitions of mixed order
or with competing orders need observables adapted to each case. Detectors that
do not presuppose the order exist. The fidelity susceptibility measures how fast
the ground state changes~\cite{ZanardiPaunkovic2006,YouLiGu2007} and has a
well-understood finite-size scaling~\cite{Albuquerque2010}, also in disordered
chains~\cite{Garnerone2009}, and networks built from correlations or mutual
information between sites change their structure at the
transition~\cite{Valdez2017}. Any new detector has to be judged against such
responses.

Topological data analysis provides tools of this kind. Persistent homology
summarizes the shape of a data set across all scales at
once~\cite{Carlsson2009,EdelsbrunnerHarer2010,ZomorodianCarlsson2005}, and it
has been used to study phase transitions in classical models, from the
topology of configuration space~\cite{Donato2016} to spin
configurations~\cite{Olsthoorn2020,Cole2021,Sale2022}, to analyze the
nonequilibrium dynamics of quantum many-body systems~\cite{Spitz2021}, and to
detect quantum phase transitions from correlations or entanglement between
sites~\cite{Tran2021,Olsthoorn2023}\rev{, or from the fidelity distance
between ground states at different values of the control
parameter~\cite{Sai2026}}. In most of these works the topological
summaries are classified, compared or fed to learning algorithms. Two
questions are less explored. Does a single topological scalar locate quantum
transitions as well as conventional responses, with a definite finite-size
behavior, and does the topology of the correlations carry information that
those responses do not?

Here we address both questions for the zeroth homology of the correlation
matrix. We treat the sites of a chain as a point cloud with the dissimilarity
$D=1-C$, where $C$ is the raw staggered spin correlation matrix of the ground
state\rev{, which builds in the antiferromagnetic pattern of the ordered
phase but requires no scalar order parameter}. The $H_0$ barcode of its
Vietoris--Rips filtration \rev{records the merge scales of the}
single-linkage hierarchy of the correlations\rev{, whose labelled merge tree
also records which sites merge}. We map \rev{the barcode} to a persistence
landscape, which has well-defined means and fluctuations~\cite{Bubenik2015},
and take the maximum of its velocity in the field as a threshold-free
pseudocritical point. \rev{The closed form of this velocity in terms of the merge
scales and its interpretation as the speed of the landscape in $L^2$ are derived in Ref.~\cite{BelchiStatuto2026}, which also shows that
in the ferromagnetic transverse-field Ising chain its maximum approaches the critical point
faster than earlier topological estimators~\cite{Olsthoorn2023}.} To judge it we compare it, on the same states and with
the same peak rule, with seven simpler responses, among them the same velocity
restricted to nearest-neighbour correlations, which isolates the role of the
merge tree, and the fidelity susceptibility.

We study the antiferromagnetic Ising chain with transverse and longitudinal fields, considering both clean and disordered systems. At zero longitudinal field, the free-fermion mapping~\cite{LiebSchultzMattis1961} provides exact correlations for large chains and many disorder realizations; at finite field, we use the density-matrix renormalization group. The model spans four regimes: the exactly known clean critical point~\cite{Pfeuty1970}, the random infinite-disorder fixed point governed by the strong-disorder renormalization group~\cite{Fisher1992,Fisher1995,IgloiLinRiegerMonthus2007}, the clean mixed-order transition line~\cite{Ovchinnikov2003,LajkoIgloi2021}, and the disorder-induced reentrant boundary near the classical multicritical point~\cite{Lajko2020}.

We find that the topological pseudocritical points locate the transition in
every regime with \rev{an accuracy comparable to that of conventional
responses, but not better than that of the best local ones}. In the
clean chain at zero field the merge tree is the chain itself, and the landscape
velocity is exactly a response of the bond energies, which explains its
finite-size scaling. In the random chain the widths of the distributions of
pseudocritical points of all estimators approach the infinite-disorder scaling
slowly. There the merge tree is no longer local, and its hierarchy
captures most of the clusters of the strong-disorder renormalization group of
each realization, including non-contiguous ones, which a hierarchy restricted
to the lattice misses\rev{, and the agreement approaches one when the disorder
is made stronger and the renormalization group becomes exact}. \rev{Correlation
clusters of many-body localized eigenstates, non-local ones included, have
been identified by cutting the same hierarchy of pairwise correlations at the
level of maximal modularity~\cite{Hemery2022}, in analogy with entanglement
clusters~\cite{Herviou2019}. Here the whole hierarchy is compared, realization
by realization, with an exact reference for the clusters.} This
information lies in the correlations of all pairs,
which other hierarchical clusterings also recover, and \rev{single linkage,
whose merge scales form the $H_0$ barcode, is its stable and}
threshold-free version\rev{~\cite{CarlssonMemoli2010}}. In a longitudinal field the
landscape velocity follows the line of mixed-order transitions and is not
distorted by the boundary wall of even open chains\rev{, a robustness that it
shares with the local responses but not with the bulk magnetization and the
fidelity susceptibility}. In odd chains the first homology
adds an interpretable boundary signal. In the reentrant regime the velocity
locates the entry into the ordered phase, an abrupt displacement of domain
walls, once the filtration is resolved by scale, and like the other rate
observables we examined it does not locate the smooth exit.

The paper is organized as follows. Section~\ref{sec:methods} introduces the
model, the numerical methods, the topological observable and the reference
estimators. Sections~\ref{sec:clean_h0}--\ref{sec:reentrant} present the four
regimes, and Sec.~\ref{sec:discussion} discusses what the topology adds and
where it does not. Numerical details are collected in
Appendix~\ref{app:numerics}, and the Supplemental Material contains \rev{the velocity without landscape
weights, further comparisons with the renormalization-group clusters, the first
homology in a longitudinal field and further analyses of the reentrant regime}.

\section{Model and methods}
\label{sec:methods}

\subsection{Model}
\label{sec:model}

We study the antiferromagnetic Ising chain in transverse and longitudinal
fields,
\begin{equation}
  \hat H
  =\sum_{i=1}^{L-1} J_i\,\sigma^z_i\sigma^z_{i+1}
  -\sum_{i=1}^{L}\Gamma_i\,\sigma^x_i
  -h\sum_{i=1}^{L}\sigma^z_i ,
  \label{eq:hamiltonian}
\end{equation}
where $\sigma^{x,z}_i$ are Pauli operators, $J_i>0$ are couplings,
$\Gamma_i>0$ are transverse fields, and $h\ge 0$ is a homogeneous
longitudinal field. We use open boundary conditions (OBC), except
for the periodic-chain comparison in Sec.~\ref{sec:reentrant}.
In the clean classical limit ($\Gamma=0$), OBC induce a parity
effect for $J<h<2J$, where field-aligned end spins are compatible with
N\'eel order for odd $L$, but force a domain wall for even $L$.
We therefore use odd $L$ wherever boundary-induced walls could
affect the results (Secs.~\ref{sec:clean_frontier}
and~\ref{sec:reentrant}).

For the clean chain, $J_i=J=1$ and $\Gamma_i=\Gamma$, with
$g=\Gamma/J$. At $h=0$, the transverse-field Ising critical
point lies at $g_c=1$~\cite{Pfeuty1970}. For $h>0$, the
antiferromagnetic phase is bounded by a line of mixed-order
transitions terminating at the classical multicritical point
$(h,\Gamma)=(2J,0)$. Along this line, the correlation length
diverges with Ising exponents~\cite{Ovchinnikov2003,LajkoIgloi2021},
while the bulk correlation function jumps~\cite{LajkoIgloi2021}.

For the random chain, following Ref.~\cite{Lajko2020}, we draw
$J_i=1+u_i\in[1,2)$ and
$\Gamma_i=\Gamma_0(1+v_i)\in[\Gamma_0,2\Gamma_0)$,
where $u_i$ and $v_i$ are independent uniform variables on $[0,1)$.
We then fix the arithmetic means of each realization through
\begin{equation}
  J_i\to J_i-\overline{J}+\tfrac32,
  \qquad
  \Gamma_i\to\Gamma_i-\overline{\Gamma}+\tfrac32\Gamma_0 ,
  \label{eq:fixed_mean}
\end{equation}
where bars denote site averages. This transformed ensemble, used
in the density-matrix renormalization group (DMRG) calculations
of Ref.~\cite{Lajko2020}, reduces finite-size sample fluctuations
without changing thermodynamic disorder averages. We keep $u_i$
and $v_i$ fixed throughout each $\Gamma_0$ scan, tracking the
same realization as the field varies. At $h=0$, the critical
point is $\Gamma_{0,c}=1$, governed by an infinite-disorder
fixed point (IDFP)~\cite{Fisher1992,Fisher1995,IgloiMonthus2005}.

\subsection{Ground states}
\label{sec:dmrg}

At $h=0$, the exact free-fermion mapping~\cite{LiebSchultzMattis1961}
provides full correlation matrices for clean and random chains
of several hundred sites, with up to a thousand realizations
per size (Secs.~\ref{sec:clean_h0} and~\ref{sec:random_h0}).
For $h>0$, we use finite-size DMRG~\cite{White1992}
implemented in TeNPy~\cite{Hauschild2018}.
We scan the field in both directions, retaining the
lower-energy branch and flagging points with branch
disagreements or saturated bond dimensions.
The protocol, flag treatment, and benchmark against
the exact $h=0$ solution up to $L=128$ are detailed
in Appendix~\ref{app:numerics}.

\subsection{Correlation matrix and distance}
\label{sec:correlations}

\begin{figure*}[t]
  \centering
  \includegraphics[width=0.95\textwidth]{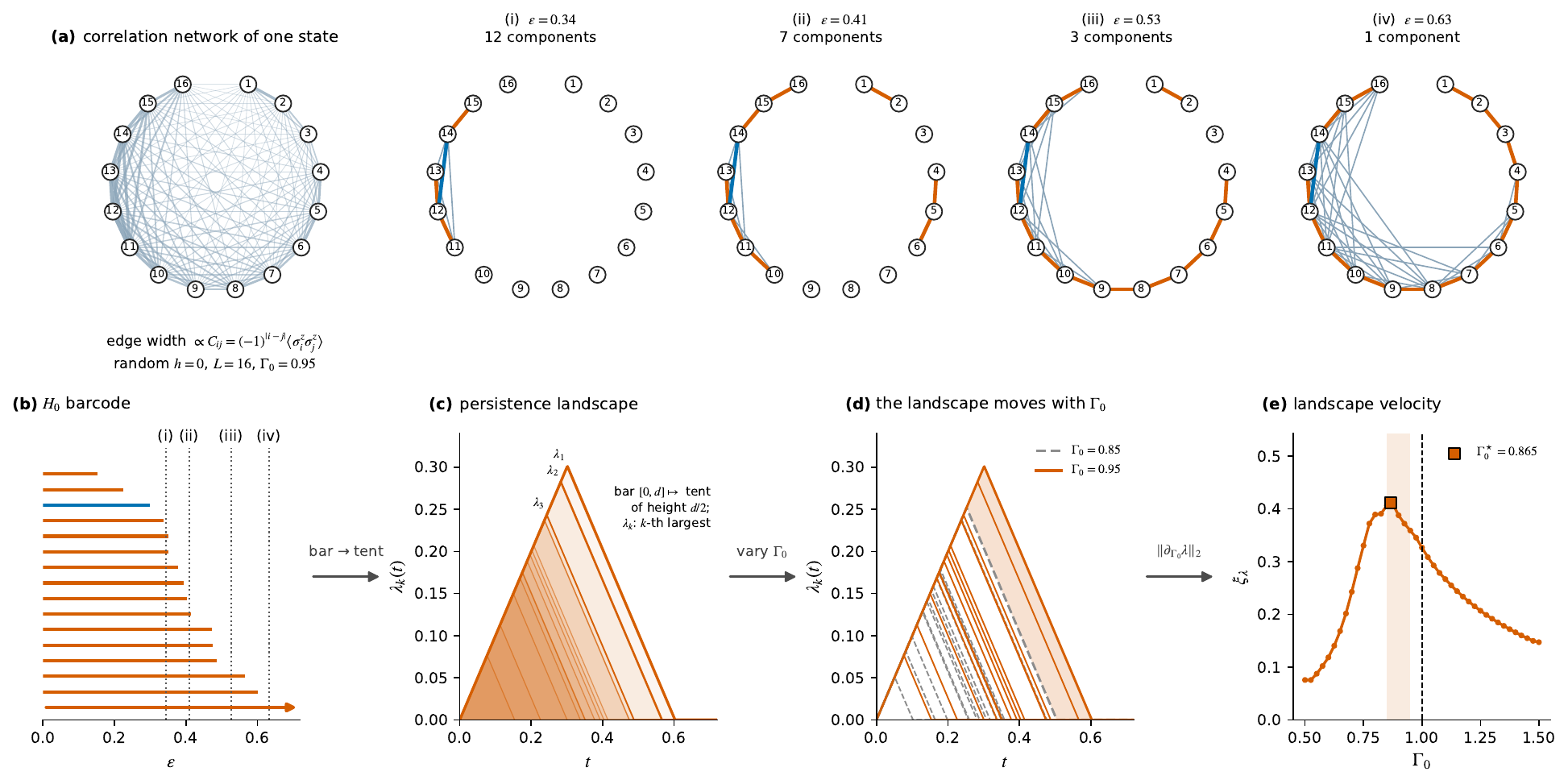}
\caption{\textbf{From a quantum state to the landscape velocity}
  for a random-chain realization ($h=0$, $L=16$, $\Gamma_0=0.95$).
  (a) Correlation network with edge widths proportional to
  $C_{ij}$, Eq.~\eqref{eq:cstag}. Snapshots (i)--(iv) show
  the filtration $D=1-C$, with edges entering at
  $D_{ij}\le\varepsilon$. Merging edges connect nearest
  neighbours (orange) or sites with $|i-j|\ge2$ (blue).
  (b) $H_0$ barcode with matching colours. Each merge at
  $\varepsilon=d$ ends a bar $[0,d)$, while the arrow marks
  the surviving component. Dotted lines indicate the snapshots.
  (c) Persistence landscape, where each bar becomes a tent
  of height $d/2$ and $\lambda_k$ is the $k$-th largest tent.
  (d) Landscapes at $\Gamma_0=0.85$ \rev{and $0.95$},
  with shading showing the change in $\lambda_1$.
  (e) Landscape velocity $\xi_\lambda$, Eq.~\eqref{eq:landscape_velocity}. The shaded band
  marks the interval in (d), and the square identifies
  the pseudocritical point at the maximum.}
  \label{fig:schematic}
\end{figure*}

From each ground state we construct the raw staggered
correlation matrix and its associated dissimilarity,
\begin{equation}
  C_{ij}=(-1)^{|i-j|}\langle\sigma^z_i\sigma^z_j\rangle ,
  \qquad
  D_{ij}=1-C_{ij} ,
  \label{eq:cstag}
\end{equation}
with $C_{ii}=1$ and $D_{ii}=0$. Unlike the connected correlator, $C$ retains the squared staggered magnetization at large separations in the antiferromagnetic phase and the field-induced uniform contribution for $h>0$, capturing their competition.
The symmetric dissimilarity $D\in[0,2]$ assigns values
below, equal to, or above $1$ to positively correlated,
uncorrelated, or negatively correlated pairs, respectively.
It need not satisfy the triangle inequality, since the
filtration only requires symmetry.

\subsection{Persistent homology and landscape velocity}
\label{sec:landscapes}

Figure~\ref{fig:schematic} illustrates the construction for a
random-chain realization ($h=0$, $L=16$, $\Gamma_0=0.95$).
We treat the sites as a point cloud with dissimilarity $D$
and compute the persistent homology of its Vietoris--Rips
filtration~\cite{Carlsson2009,EdelsbrunnerHarer2010,
ZomorodianCarlsson2005}, connecting sites $i$ and $j$
when $D_{ij}\le\varepsilon$. The zeroth homology $H_0$
tracks the merging of connected components as
$\varepsilon$ increases from zero, initially with one
component per site. Figure~\ref{fig:schematic}(a) shows
the correlation network, with edge widths proportional
to $C_{ij}$, and snapshots containing 12, 7, 3 and 1
components at $\varepsilon=0.34, 0.41, 0.53$ and $0.63$.
Merging edges connect mostly nearest neighbours (orange), but also non-neighbouring sites (blue), showing that 
$H_0$
follows the correlation hierarchy rather than being restricted to the nearest-neighbour connectivity of the underlying one-dimensional lattice.

Each merge at scale $d_a$ produces a finite bar $[0,d_a)$
in the $H_0$ barcode, see Fig.~\ref{fig:schematic}(b).
The $L-1$ finite bars span $d=0.15$ to $0.60$ in this
example, while the infinite bar (arrow), which we discard,
represents the surviving component. At the four dotted
snapshot scales, the number of surviving bars, including
the infinite one, equals the number of components in
Fig.~\ref{fig:schematic}(a). 
The merging scales are the edge weights of the minimum
spanning tree of $D$, equivalently the maximum spanning
tree of $C$, and form the $H_0$ barcode of the
single-linkage hierarchy~\cite{Carlsson2009}.
\rev{The barcode does not identify which sites merge, as
different hierarchies can share the same barcode and
landscape. The SDRG comparison in
Sec.~\ref{sec:random_clusters} therefore uses the labelled
merge tree returned by Kruskal's algorithm, with a fixed
tie-breaking rule for equal edge weights. 
For the Vietoris--Rips filtration used here, $H_0$
depends only on the filtered graph, in which sites
$i$ and $j$ are joined by an edge whenever
$D_{ij}\leq\varepsilon$. Higher-dimensional simplices do not affect connected
components or their merge scales. Thus the labelled
merge hierarchy can be computed directly from this
weighted graph, without constructing the full
Vietoris--Rips filtration~\cite{BelchiStatuto2026}.}

We use persistence landscapes~\cite{Bubenik2015}, Fig.~\ref{fig:schematic}(c), to compare barcodes across fields.
Each finite bar $[0,d_a(g))$ defines a tent
\[
\phi_a(g,t)
=
\max\bigl\{0,\min\{t,d_a(g)-t\}\bigr\},
\]
of height $d_a(g)/2$. In the notation of the companion
mathematics paper \cite{BelchiStatuto2026}, this is
$\Lambda_{(0,d_a(g))}(t)$. Here the control parameter is
$g=\Gamma/J$ for clean chains and $g=\Gamma_0$ for random
ones.

The $k$-th landscape layer is
\begin{equation}
  \lambda_k(g,t)=
  \underset{a=1,\ldots,L-1}{k\text{-}\max}\,
  \phi_a(g,t),
  \qquad k=1,\ldots,L-1,
  \label{eq:landscape}
\end{equation}
where $k\text{-}\max$ denotes the $k$-th largest of the
$L-1$ tent values at fixed $(g,t)$, counting repeated
values separately.
Thus, $\lambda_1$ captures the longest bars, while
other layers represent shorter ones.

Unlike barcodes, whose Fr\'echet means need not be
unique~\cite{Mileyko2011,Turner2014}, landscapes belong
to the Hilbert space $L^2$, allowing differences,
derivatives and averages, with sample means obeying
the law of large numbers and the central limit
theorem~\cite{Bubenik2015}. This representation
consistently weights $H_0$ bars and extends to averages
and higher homology. 

The landscape representation is also stable. If
\[
\max_{i,j}|C_{ij}-\widetilde C_{ij}|\leq\eta,
\]
then $\max_{i,j}|D_{ij}-\widetilde D_{ij}|\leq\eta$.
The bottleneck distance between the corresponding
barcodes is at most $\eta$
~\cite{CohenSteiner2007,ChazalProximity2009}, and the
uniform distance between their landscapes satisfies
\[
\|\lambda-\widetilde\lambda\|_\infty
:=
\sup_{k,t}|\lambda_k(t)-\widetilde\lambda_k(t)|
\leq\eta
\]
~\cite{Bubenik2015}. These bounds apply to the
Vietoris--Rips filtrations on the same labelled sites
and do not require $D$ to satisfy the triangle inequality.

For $H_0$, the merge scales, ordered in the same way
for both inputs, satisfy
$|d_k-\widetilde d_k|\leq\eta$
~\cite{BelchiStatuto2026}. Consequently, if each correlation $C_{ij}(g)$ depends
continuously on the control parameter $g$, then the
landscape also depends continuously on $g$ in the
uniform norm, including at crossings of merge scales.
The labelled merge hierarchy is also stable when
represented by its cophenetic distances
\[
u_D(i,j)
=
\min_{\gamma:i\to j}\max_{e\in\gamma}D_e,
\]
where $\gamma$ ranges over graph paths joining $i$ to $j$.
Here $u_D(i,j)$ is the first filtration scale at which
$i$ and $j$ belong to the same connected component, and
\[
\max_{i,j}
|u_D(i,j)-u_{\widetilde D}(i,j)|
\leq\eta
\]
~\cite{CarlssonMemoli2010}.

These stability bounds control the landscapes themselves,
but do not by themselves bound errors in their field
derivatives.

Figure~\ref{fig:schematic}(d) illustrates the landscape
change between $\Gamma_0=0.85$ and $0.95$, with shading
showing the change in $\lambda_1$.
To quantify the rate of this change with respect to $g$,
we define the persistence-landscape velocity as
\begin{equation}
  \xi_\lambda(g,L)
  =\left[\sum_{k=1}^{L-1}\int_0^\infty
  \bigl|\partial_g\lambda_k(g,t)\bigr|^2\,\mathrm{d}t\right]^{1/2}.
  \label{eq:landscape_velocity}
\end{equation}
\rev{Where differentiable, $\xi_\lambda$
is the landscape speed in $L^2$~\cite{BelchiStatuto2026},
analogous to the ground-state speed in the Fubini--Study
metric given by $\sqrt{\chi_F}$ (Sec.~\ref{sec:peaks}).}
For $H_0$, all bars start at zero, so $\lambda_k$ is
the tent of the $k$-th longest bar, whose field dependence
lies in its descending side. Equation~\eqref{eq:landscape_velocity}
then reduces to~\cite{BelchiStatuto2026}
\begin{equation}
  \xi_\lambda^2=\frac{1}{2}\sum_{k=1}^{L-1}d_{k}
  \bigl(\partial_g d_{k}\bigr)^2 ,
  \label{eq:velocity_deaths}
\end{equation}
where $d_{k}$ is the $k$-th largest merge scale.
Thus, the velocity weights changes in merge scales by
their lengths, emphasizing the weakest links in the
correlation hierarchy. \rev{For correlations Lipschitz
in $g$, Eq.~\eqref{eq:velocity_deaths} holds almost
everywhere, with one-sided velocities at merge-scale
crossings~\cite{BelchiStatuto2026}. The landscape
definition also extends to higher homology and to
sample averages, whose velocity can be obtained from
the individual barcodes even though the mean landscape
need not represent a single barcode~\cite{BelchiStatuto2026}.}

We compute field derivatives by finite differences
and landscape integrals by the trapezoidal rule
(Appendix~\ref{app:numerics}). The velocity uses all
bars without persistence thresholds or selecting a
chain region or distance. Although the staggered
correlator from Eq.~\eqref{eq:cstag} encodes antiferromagnetic
order, the maxima persist with absolute and unstaggered
correlations, and with connected correlations in chains
without boundary walls (Appendix~\ref{app:numerics}).
Figure~\ref{fig:schematic}(e) shows a maximum at
$\Gamma_0=0.865$, defining the pseudocritical point
$g^\star_\lambda(L)=\arg\max_g\xi_\lambda(g,L)$.
A peak that grows and sharpens with $L$ supports
critical finite-size scaling but does not exclude
a crossover (Sec.~\ref{sec:clean_frontier}).

For each disorder realization $\alpha$, we compute
$\lambda^{(\alpha)}$, $\xi_\lambda^{(\alpha)}$ and
$\Gamma^\star_{0,\alpha}$ separately, retaining the
labelled merge tree for the SDRG comparison.

We also compute the first homology $H_1$ of the same
Vietoris--Rips filtration using GUDHI~\cite{Maria2014}
with $\mathbb{Z}_2$ coefficients. As a complement to
$H_0$, we use the number $N_1(\tau)$ of $H_1$ intervals
with lifetime exceeding a threshold $\tau$.

\subsection{Pseudocritical points and physical reference}
\label{sec:peaks}

We locate all topological and physical maxima by refining
the discrete peak with a three-point parabolic fit.
Curves are plotted on the original grid without smoothing
or interpolation. Our physical reference is the squared
staggered magnetization of the central half of the chain,
\begin{equation}
  m^2_{s,\mathcal B}=\frac{1}{|\mathcal B|^2}
  \sum_{i,j\in\mathcal B}C_{ij},
  \qquad
  \mathcal B=\{\lfloor L/4\rfloor+1,\ldots,
  \lfloor 3L/4\rfloor\},
  \label{eq:bulk_ms2}
\end{equation}
whose response $-\partial_g m^2_{s,\mathcal B}$ defines
the pseudocritical point $g^\star_m(L)$ at its maximum.
For disordered chains, we evaluate it for each realization.
The reentrant regime uses a different reference
(Sec.~\ref{sec:reentrant}).

To assess the contribution of topology, we compare
$\xi_\lambda$ with simpler responses of the same states.
These include the landscape velocity of nearest-neighbour
distances $1-C_{i,i+1}$, which isolates the role of the
merge tree, the derivative of the mean nearest-neighbour
correlation, $\|\partial_g C\|_F/L$, the second derivative of the energy density e, $-\partial_g^2e$,
the derivative of the half-chain entanglement entropy
$S_{L/2}$, and the fidelity susceptibility
$\chi_F=-2\ln|\langle\psi(g)|\psi(g+\delta g)\rangle|/
\delta g^2$~\cite{ZanardiPaunkovic2006,YouLiGu2007}.
The latter requires no correlation matrix
\rev{and its underlying fidelity distance has also been
used in persistent homology~\cite{Sai2026}}.
Results are summarized in Table~\ref{tab:benchmark}
of Appendix~\ref{app:benchmark}.

\section{Clean chain at zero longitudinal field}
\label{sec:clean_h0}
% Fig. 2 (V1, paper_figures/fig_clean_h0): exact free fermions. Numbers from
% paper_figures/fig_clean_h0/output/fig_clean_h0_summary.json (Cl* macros).

We first test the landscape velocity in the exactly
solvable clean chain at $h=0$. Rotating every second
spin by $\pi$ about the $x$ axis maps the antiferromagnet
and its staggered correlator $C_{ij}$, Eq.~\eqref{eq:cstag}, onto their ferromagnetic
counterparts. The critical point is $g_c=1$, with
exponents $\nu=z=1$~\cite{Pfeuty1970}.
We compute exact ground states for $L=16$ to $512$,
doubling the size at each step, on a uniform grid
$g\in[0.5,1.5]$ with spacing $0.005$.

\begin{figure*}[t]
  \centering
  \includegraphics[width=\textwidth]{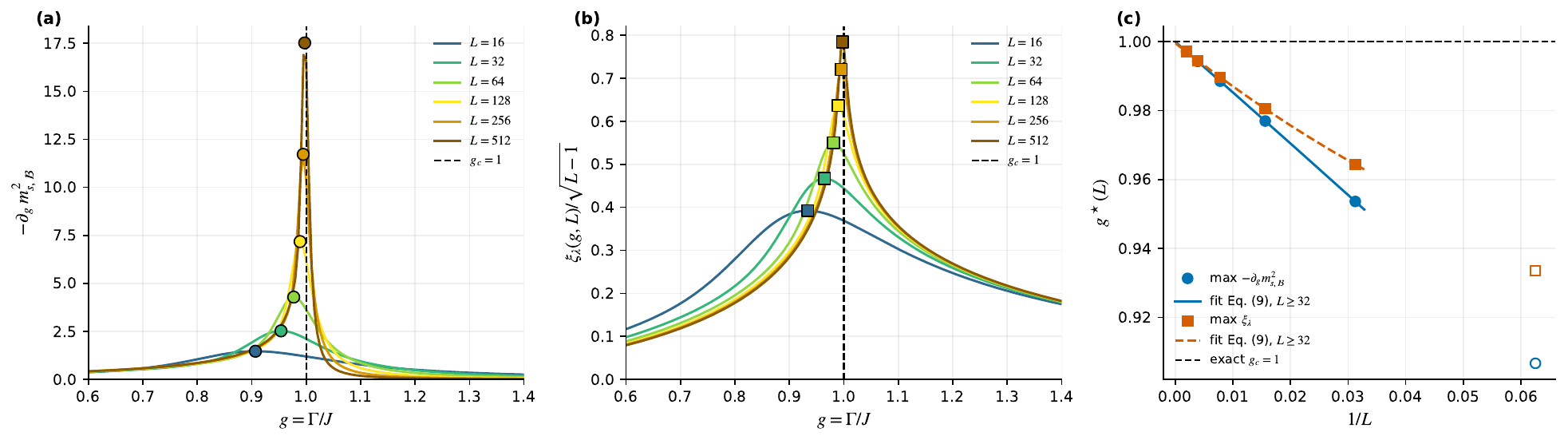}
  
\caption{\textbf{Clean chain at $h=0$ with open boundaries},
  exact ground states for $L=16$ to $512$.
  (a) Field response $-\partial_g m^2_{s,\mathcal B}$
  of the bulk staggered magnetization, Eq.~\eqref{eq:bulk_ms2}.
  (b) $H_0$ landscape velocity
  \rev{per finite bar, $\xi_\lambda(g,L)/\sqrt{L-1}$}, Eq.~\eqref{eq:landscape_velocity}.
  Symbols mark the maxima and the dashed line marks
  $g_c=1$.
  (c) Pseudocritical points $g^\star(L)$ from
  (a) (circles) and (b) (squares) versus $1/L$,
  \rev{refined off the field grid
  (Appendix~\ref{app:numerics})}.
  Lines are fits of Eq.~\eqref{eq:inverse_size_fit}
  for $L\ge32$, excluding the open symbols ($L=16$).
  Maxima follow the rule of Sec.~\ref{sec:peaks}.}

  \label{fig:clean_h0}
\end{figure*}

Figure~\ref{fig:clean_h0}(a) shows the physical response
$-\partial_g m^2_{s,\mathcal B}$ from Eq.~\eqref{eq:bulk_ms2}, whose maximum grows and
sharpens with $L$, approaching $g_c$ from below and
reaching $g=0.9971$ at $L=512$. Figure~\ref{fig:clean_h0}(b) shows the landscape velocity from Eq.~\eqref{eq:landscape_velocity},
\rev{normalized by $\sqrt{L-1}$ to compare sizes without
shifting the maxima}, computed over the entire chain.
It behaves similarly, with peak heights increasing from
$0.39$ at $L=16$ to $0.78$ at $L=512$, and the peak
approaching $g_c$ from below, reaching $g=0.9972$
at $L=512$. Away from criticality, the curves nearly
coincide because the correlation length is short and
the $H_0$ merge scales and their derivatives become
size-independent. The peak growth is thus confined
to the critical region.

To extrapolate the maxima, we use the finite-size scaling
of pseudocritical points,
\begin{equation}
  g^\star(L)-g_c\propto L^{-1/\nu},
  \label{eq:shift}
\end{equation}
\cite{FerdinandFisher1969,FisherBarber1972,Barber1983}.
Near criticality, observables scale with $L/\xi$, where
$\xi\propto|g-g_c|^{-\nu}$, so their maxima approach
$g_c$ with the shift exponent $1/\nu$. The shift
amplitude and sign depend on the observable and boundary
conditions. Using the exact $\nu=1$ with a leading
correction to scaling, we fit both sets of maxima to
\begin{equation}
  g^\star(L)=g_c+\frac{a}{L}+\frac{b}{L^2},
  \label{eq:inverse_size_fit}
\end{equation}
without weights for $L\ge32$, excluding the smallest
chain from the fits.

Figure~\ref{fig:clean_h0}(c) shows the pseudocritical
points against $1/L$\rev{, with maxima refined off the
simulation grid (Appendix~\ref{app:numerics})}.
\rev{Fitting Eq.~\eqref{eq:inverse_size_fit} gives
$g_c=0.9996$ for $\xi_\lambda$ and $1.0000$ for the
physical response, with standard errors of $10^{-4}$
and $6\times10^{-6}$. These errors are conditional on
the three-parameter ansatz fitted to five sizes.
Replacing $b/L^2$ by $c\ln L/L^2$, using
$g^\star=g_c+(a+c\ln L)/L$ or
$g^\star=g_c+aL^{-1/\nu}$, and varying the minimum
fitted size from $L=16$ to $64$ gives
$g_c\in[0.9991,1.0005]$ for $\xi_\lambda$ and
$[0.99994,1.00006]$ for the physical response.
These systematic ranges agree with the exact $g_c=1$.
For periodic chains, convergence to $g_c$ follows
from the exact solution~\cite{BelchiStatuto2026}, while for open
chains it remains numerical.}
The leading amplitudes are $a=-1.32$ for
$\xi_\lambda$ and $a=-1.46$ for the physical response.

\rev{Free-power fits yield effective exponents
$1/\nu_{\mathrm{eff}}=0.86$--$0.89$ for $\xi_\lambda$
and $1.01$ for the physical response.
With $g_c=1$ fixed, the local exponents of
$\xi_\lambda$ decrease from $0.90$ to $0.88$ before
rising to $0.95$, while those of the physical response
remain between $1.002$ and $1.010$.
A leading logarithmic correction fits the
$\xi_\lambda$ maxima for $L\ge32$ with residuals
below $10^{-5}$, $c=-0.16$ and $g_c=1.0003$,
but five sizes cannot distinguish it from an effective
power law.}
Thus, the topological pseudocritical points converge
towards the exact critical point with larger scaling
corrections than the physical response, consistent
with the logarithmic divergence of the bond-energy
response discussed below. The effective exponents
therefore quantify these corrections rather than
provide an independent estimate of $\nu$.

At $h=0$, the clean-chain construction of
Sec.~\ref{sec:landscapes} reduces to nearest-neighbour
correlations. In our data, every nonadjacent pair is
less correlated than the intervening nearest-neighbour
pairs, so the minimum spanning tree of $D$ follows the
lattice. \rev{We verified this numerically at all fields
for $L\le512$, also in periodic chains~\cite{BelchiStatuto2026}.}
The merge scales are thus $1-C_{i,i+1}$, and
Eq.~\eqref{eq:velocity_deaths} expresses $\xi_\lambda$
as a weighted norm of bond-energy derivatives.
These diverge logarithmically at $g_c$, as does the
Ising specific heat with $\alpha=0$. Accordingly, the
peak of \rev{$\xi_\lambda/\sqrt{L-1}$} grows linearly
with $\ln L$, with slope $0.116$ for $L=16$--$512$.
\rev{For periodic chains, this growth follows from
the exact solution~\cite{BelchiStatuto2026}.}
Thus, the $h=0$ case tests consistency rather than
providing additional topological information.
The merge tree departs from the lattice when
correlations are not ordered by distance, as in
random chains (Sec.~\ref{sec:random_h0}) or at finite
longitudinal field (Sec.~\ref{sec:clean_frontier}).

\section{Random chain at zero longitudinal field}
\label{sec:random_h0}
% V1. Fig. 3 (paper_figures/fig_random_h0) and Fig. 4 (paper_figures/fig_h0_clusters), exact free fermions.
% Numbers: fig_random_h0_summary.json, results/review_v1/free-fermion-h0-v1/analysis/summary.json,
% results/review_v1/sdrg-vs-h0-v1/{summary,hierarchy_summary}.json (Rd*, Sd*, Sch* macros).

In the clean chain, the maximum of $\xi_\lambda$
follows the finite-size scaling of the transition.
With quenched disorder, each realization has its own
pseudocritical point. We ask (i) whether their
distribution follows the predictions of the disordered
fixed point and \rev{(ii)} whether the merge tree,
no longer constrained to the lattice, contains
additional information.

At $h=0$, the random transverse-field Ising chain
has a self-dual critical point at $\Gamma_0=1$ for
the box distributions of Sec.~\ref{sec:model}.
Its infinite-disorder fixed point (IDFP) has
correlation-length exponents $\nu=2$ for average
correlations and $\nu_{\mathrm{typ}}=1$ for typical
ones~\cite{Fisher1992,Fisher1995}.
The distribution of sample-dependent pseudocritical
points has a width governed by the disordered
fixed point~\cite{AharonyHarris1996,WisemanDomany1998}.
For this model, widths obtained from entanglement
entropy, surface magnetization and soft-mode energies
scale as $L^{-1/\nu}$ with $\nu=2$
~\cite{IgloiLinRiegerMonthus2007}.
Since the mean shift is smaller and depends on the
definition~\cite{IgloiLinRiegerMonthus2007}, we use
the width to test the fixed point, noting its
sensitivity to the disorder ensemble~\cite{Monthus2004}.
We compute exact ground states for $1000$
realizations per size at $L=16$--$128$ and $500$
at $L=256$, on a grid of spacing $0.005$ over
$\Gamma_0\in[0.5,1.5]$, restricted to $[0.7,1.3]$
for $L=256$.

\subsection{Pseudocritical points of single chains}
\label{sec:random_peaks}

\begin{figure}[t]
  \centering
  \includegraphics[width=0.95\columnwidth]{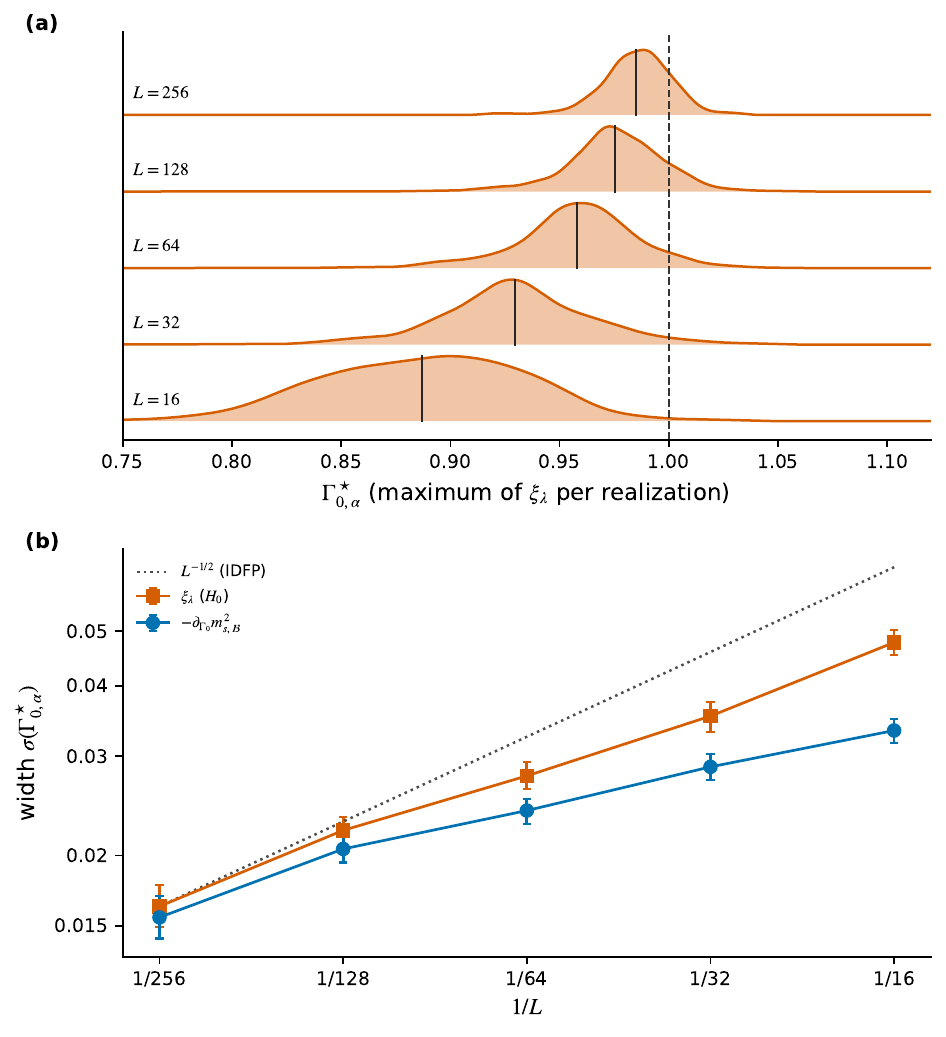}
  
\caption{\textbf{Random chain at $h=0$ with open
boundaries and fixed-mean disorder}, exact ground
states. (a) Distribution of topological pseudocritical
points $\Gamma^\star_{0,\alpha}$ from individual
$\xi_\lambda$ maxima for $L=16$--$256$
($1000$ realizations, $500$ at $L=256$).
Vertical lines mark the means and the dashed line
the critical point $\Gamma_{0,c}=1$.
(b) Widths of topological (squares) and physical
(circles) pseudocritical-point distributions versus
$1/L$ on logarithmic axes, with 95\% bootstrap
intervals. The dotted line indicates the
infinite-disorder scaling $L^{-1/2}$.}
  \label{fig:random_h0}
\end{figure}

Figure~\ref{fig:random_h0}(a) shows, for each size, the distribution of the
maxima $\Gamma^\star_{0,\alpha}$ of the velocities $\xi^{(\alpha)}_\lambda$ of
the individual realizations $\alpha$. The distribution narrows and moves
towards $\Gamma_{0,c}=1$ as $L$ grows, and for $L=256$ its mean is
$0.985$, against $0.991$ for the maxima of the
physical response $-\partial_{\Gamma_0}m^2_{s,\mathcal B}$ of the same
realizations.

Figure~\ref{fig:random_h0}(b) shows the distribution
widths, decreasing from $0.048$ to $0.016$ for
$\xi_\lambda$ and from $0.033$ to $0.016$ for the
physical estimator between $L=16$ and $256$.
Power-law fits $\sigma\propto L^{-p_\sigma}$
over all sizes, with 95\% intervals from $2000$
bootstrap resamplings~\cite{Efron1979}, give
$p_\sigma=0.38$ $[0.35,0.41]$ and $0.27$
$[0.24,0.30]$, respectively. Both remain below
the IDFP value $1/\nu=1/2$, although the local
exponents increase to $0.45$ and $0.40$ between
the largest sizes. Similar behavior occurs for
the other estimators in Table~\ref{tab:benchmark}
of Appendix~\ref{app:benchmark}.
\rev{These effective exponents suggest a slow
preasymptotic approach but do not establish
IDFP scaling.} Such corrections are expected
for moderate box disorder, which strengthens
under renormalization~\cite{Fisher1995,IgloiMonthus2005}.

\rev{The fixed-mean ensemble constrains arithmetic
means, while sample criticality depends on
$\delta_\alpha=\overline{\ln\Gamma}-\overline{\ln J}$,
which still fluctuates as $L^{-1/2}$ with reduced
amplitude. These fluctuations are much smaller
than the pseudocritical-point widths, which depend
on the spatial disorder configuration.
The expected $\sigma\propto L^{-1/2}$ follows instead from finite-size scaling at the IDFP \cite{IgloiLinRiegerMonthus2007}, and the fixed-mean ensemble should change its prefactor and corrections rather than its exponent
~\cite{Monthus2004,Pazmandi1997}.
To test ensemble effects, we also use the same
random numbers without the shift of
Eq.~\eqref{eq:fixed_mean} ($500$ realizations,
$L=16$--$128$). The resulting widths are
$1.5$--$1.7$ times larger for $\xi_\lambda$
and $1.6$--$2.2$ times larger for the physical
response. Their fitted exponents are
$p_\sigma=0.41$ $[0.37,0.45]$ and
$0.36$ $[0.32,0.40]$, compared with $0.35$
and $0.22$ for the fixed-mean ensemble,
but remain below $1/2$.}
The larger global exponent of $\xi_\lambda$
reflects its greater width at small $L$ rather
than faster convergence. \rev{Both pseudocritical
distributions therefore have comparable widths
and show similar preasymptotic behavior.}

\rev{To compare estimators sample by sample, we
correlate their pseudocritical points with references
requiring no ground state. At $L=256$, the sample
self-dual point, defined by
$\overline{\ln\Gamma}=\overline{\ln J}$, has a
standard deviation only $0.10$ times that of the
topological pseudocritical points. Both estimators
correlate weakly with it, with Pearson coefficients
of $0.09$--$0.12$ for $\xi_\lambda$ and
$0.11$--$0.17$ for the physical response across
sizes. Thus, spatial disorder matters more than
its global balance.

\begin{figure*}[t]
  \centering
  \includegraphics[width=0.95\textwidth]{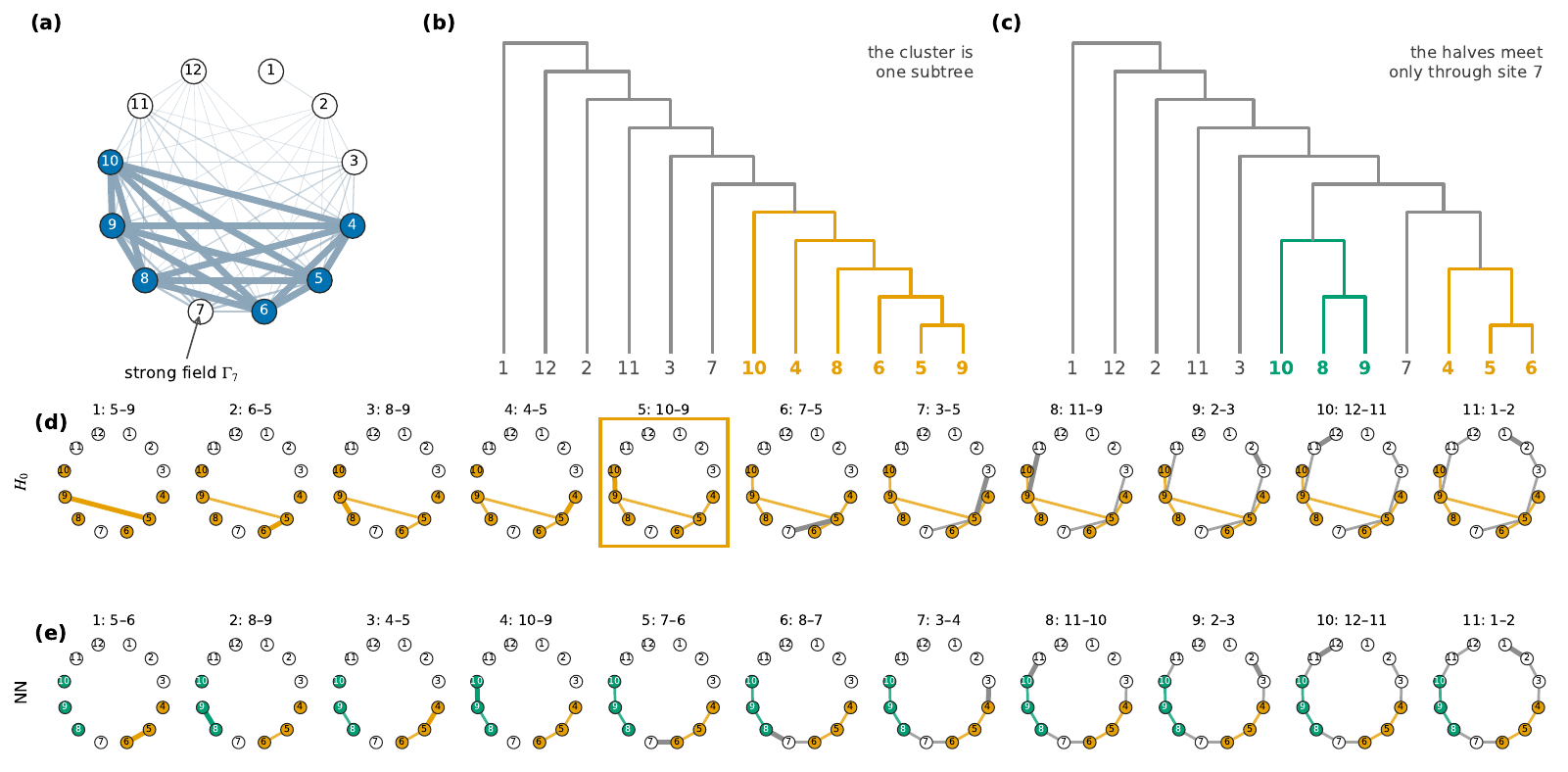}
  \caption{\textbf{How the $H_0$ hierarchy detects a cluster of the strong-disorder
  RG}, for an illustrative open chain of $L=12$ sites at $h=0$ with chosen
  couplings and its exact ground state. Site 7 carries a strong transverse
  field, and the RG joins sites 4 to 6 and 8 to 10 into one non-contiguous
  cluster (filled circles). (a)~Correlation network, drawn as in
  Fig.~\ref{fig:schematic}. (b)~$H_0$ hierarchy, the single-linkage tree of
  $D=1-C$. The cluster is one subtree (yellow). (c)~Hierarchy restricted to
  nearest-neighbour links. The two halves (yellow, green) merge only through
  site 7. Branch heights show the merge order. (d), (e)~Tree edges after each
  merge $k$ of (b) and (c), with the newest edge thicker and edges outside the
  cluster in grey. In (d) the cluster is complete after merge 5 (framed)
  without site 7.}
  \label{fig:h0_clusters}
\end{figure*}

Fixed means induce a strong anticorrelation
($-0.98$) between the self-dual points of the
central half $\mathcal B$ and the end quarters.
The physical response in $\mathcal B$ correlates
with its local self-dual point, decreasing from
$0.84$ at $L=16$ to $0.64$ at $L=256$.
The end-sensitive surface-magnetization
pseudocritical point~\cite{Peschel1984} is instead
anticorrelated with it ($-0.44$ to $-0.32$).

By contrast, $\xi_\lambda$ probes the whole chain
and correlates most strongly with global responses,
with coefficients of $0.68$--$0.81$ for fidelity
susceptibility, $0.62$--$0.78$ for the energy
second derivative and $0.68$--$0.75$ for
$\|\partial_g C\|_F$, compared with
$0.44$--$0.56$ for the response in $\mathcal B$.
Thus, the topological estimator tracks individual
realizations similarly to other global responses.}

\subsection{\texorpdfstring{$H_0$}{H0} hierarchy and the clusters of the strong-disorder RG}
\label{sec:random_clusters}

Unlike in the clean chain, the random-chain merge
tree need not follow the lattice. Near $\Gamma_0=1$,
non-nearest-neighbour merges increase from $1\%$
at $L=16$ to $13\%$ at $L=256$.
The strong-disorder renormalization group
(SDRG)~\cite{Fisher1992,Fisher1995,IgloiMonthus2005}
explains their origin by successively decimating
the largest \rev{Hamiltonian term, either a bond
or a transverse field}. A strong bond $J_i$
merges clusters, generating an effective field
$\tilde\Gamma=\Gamma_A\Gamma_B/J_i$, while a
strong field $\Gamma_i$ freezes a cluster
\rev{along the $x$ direction} and couples its
neighbours through
$\tilde J=J_{i-1}J_i/\Gamma_i$.
Correlated clusters can therefore span sites
removed by field decimations and need not be
contiguous. SDRG becomes asymptotically exact
at the IDFP~\cite{Fisher1995}.

Figure~\ref{fig:h0_clusters} illustrates this
structure for a short chain with chosen couplings.
After decimating site 7, SDRG joins sites 4--6
and 8--10 into one cluster. The ground-state
correlations remain strong across site 7,
$C_{68}=0.97$, compared with $C_{67}=0.50$
involving site 7, as shown in
Fig.~\ref{fig:h0_clusters}(a).
The single-linkage tree of $D$ connects the
two halves directly, starting with the
non-nearest-neighbour pair 5--9, and recovers
the six-site cluster as a subtree after five
merges, as shown in Fig.~\ref{fig:h0_clusters}(b)
and (d). Restricting links to nearest neighbours
forces the halves to connect through site 7,
preventing the cluster from appearing independently,
as shown in Fig.~\ref{fig:h0_clusters}(c) and (e).

We ran SDRG at $\Gamma_0=1$ for each realization
until all sites were decimated, assigning each site
to the cluster removed by its field and excluding
single-site clusters. \rev{This operational definition
retains clusters at all decimation scales.}
For $300$ realizations per size, from $L=32$ to
$512$, we compared these clusters with the
\rev{labelled merge tree of the} $H_0$ hierarchy
from exact correlations. \rev{The barcode alone
contains only merge scales
(Sec.~\ref{sec:landscapes}).}
About $59\%$ of sites belong to clusters of at
least two sites, $52\%$ of which are pairs.
At $L=512$, each realization contains on average
$10.5$ non-contiguous clusters of at least three
sites, and the largest cluster is non-contiguous
in at least $95\%$ of realizations for $L\ge64$.

We quantify agreement using dendrogram
purity~\cite{HellerGhahramani2005}, averaging
over pairs within each SDRG cluster the fraction
of leaves in their smallest common subtree
belonging to that cluster. A cluster of $n$ sites
thus contributes $n(n-1)/2$ pairs, and purity
equals one when every cluster forms a subtree.
Equal cluster weighting instead gives
$0.87$--$0.92$ at $L=512$ for all hierarchies,
including nearest-neighbour, because contiguous
pairs dominate. Pair weighting emphasizes
larger clusters, where the hierarchies differ.

\begin{figure}[t]
  \centering
  \includegraphics[width=0.95\columnwidth]{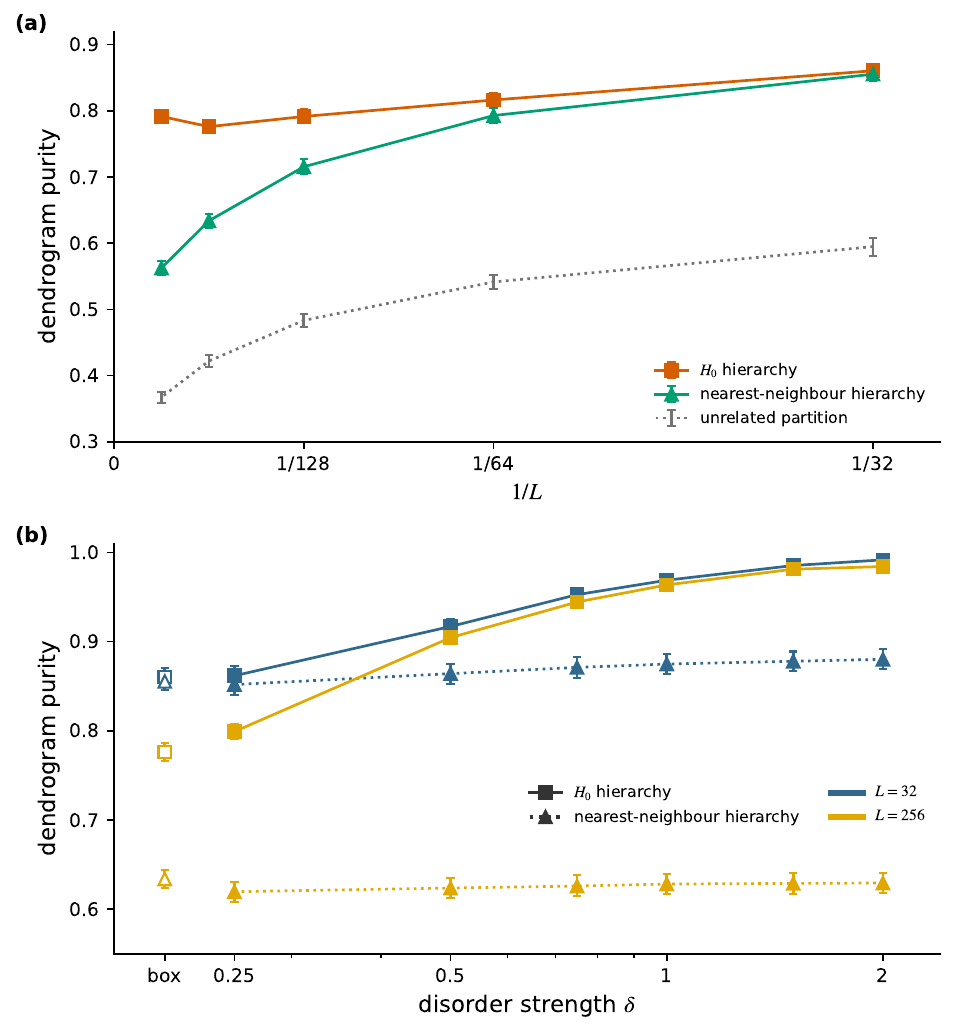}
\caption{\rev{\textbf{Agreement of the $H_0$ hierarchy
with strong-disorder RG clusters} at $h=0$ and
$\Gamma_0=1$, measured by dendrogram purity.
(a) Box disorder of Fig.~\ref{fig:random_h0}
versus $1/L$, comparing $H_0$ (squares),
nearest-neighbour (triangles) and unrelated
partitions (dotted).
(b) Power-law disorder versus strength $\delta$
for $L=32$ and $256$, comparing $H_0$ (solid)
and nearest-neighbour (dotted) hierarchies.
Open symbols show the box-disorder values
at their standard deviation of $\ln J$.
Each point uses $300$ realizations, with
95\% bootstrap intervals.}}
  \label{fig:sdrg_purity}
\end{figure}

\rev{Figure~\ref{fig:sdrg_purity}(a)} shows that
$H_0$ purity initially decreases before
stabilizing at $0.78$--$0.86$, while
nearest-neighbour purity falls from $0.86$
at $L=32$ to $0.56$ at $L=512$.
Using clusters from another realization gives
$0.59$ and $0.37$, respectively.
\rev{The paired purity advantage of $H_0$
increases from $0.005$ $[0.003,0.007]$ at
$L=32$ to $0.23$ $[0.22,0.24]$ at $L=512$,
with 95\% bootstrap intervals. At $L=512$,
$H_0$ is purer in all $300$ realizations.}
This advantage comes from non-contiguous clusters
of at least three sites, whose purity at
$L=512$ is $0.79$ for $H_0$ versus $0.55$
for nearest-neighbour. The best matching
subtree likewise achieves a Jaccard index
of $0.76$, compared with $0.66$ for
nearest-neighbour and $0.43$ for the
unrelated partition.

The cluster information comes from the full pairwise
correlations rather than single linkage itself.
Average linkage on $D$ gives \rev{nearly} identical
purity to $H_0$ at $L=512$ ($0.79$ for both), while
complete linkage gives $0.69$.
\rev{Paired differences between single and average
linkage remain below $0.007$ at every size.
A single threshold performs worse. Although the
components of $C_{ij}>c$ are subtrees of the $H_0$
hierarchy, optimizing $c$ for each realization
gives a Jaccard index of only $0.64$ at $L=512$,
against $0.76$ for the best matching subtrees,
because clusters emerge at different scales.
Hémery \textit{et al.} used a maximal-modularity
cut of the same hierarchy to identify correlation
clusters in many-body localized
eigenstates~\cite{Hemery2022}.
Here we compare the full hierarchy with the SDRG
clusters of each realization. At $L=512$, the
maximal-modularity cut yields about ten blocks,
compared with $257$ SDRG clusters, and a Jaccard
index of only $0.22$. Selecting persistent
subtrees with HDBSCAN~\cite{Campello2013,Pedregosa2011},
without using the SDRG reference, improves this
to $0.43$, outperforming the modularity cut
in $289$ of $300$ realizations, but remains
below the $0.76$ available in the full hierarchy
(Sec.~SII\,C of the Supplemental Material).}

Thus, the full correlation hierarchy captures
much of the SDRG cluster structure, including
non-contiguous clusters, without running SDRG.
\rev{The comparable performance of average linkage
shows that recovery is not specific to single
linkage. However, single linkage uniquely satisfies
the axioms of Ref.~\cite{CarlssonMemoli2010},
has cophenetic distances stable against correlation
errors (Sec.~\ref{sec:landscapes}), unlike the
tie-breaking dependence of average and complete
linkage, and yields the $H_0$ barcode, linking
pseudocritical points and cluster structure.}
Agreement remains partial at moderate disorder,
where SDRG is not yet exact, but saturates with
$L$, unlike the nearest-neighbour hierarchy.

\rev{We examined stronger disorder using
$J_i=u_i^\delta$ and
$\Gamma_i=\Gamma_0v_i^\delta$, with $u_i,v_i$
uniform in $(0,1]$. The logarithmic couplings
and fields have standard deviation $\delta$,
and SDRG becomes exact as $\delta$
increases~\cite{Fisher1995,IgloiMonthus2005}.
Each sample is shifted to the ensemble mean
logarithms, analogously to
Eq.~\eqref{eq:fixed_mean}, placing it at its
self-dual point for $\Gamma_0=1$.
For comparison, the box distribution has
$\mathrm{std}(\ln J)\simeq0.2$.
Since all couplings and fields share the same
power $\delta$, the SDRG decimation rules
produce identical clusters for a given
realization at every $\delta$, while the exact
ground state changes.

For $300$ realizations per size at
$L=32$--$256$ and $\delta=0.25$--$2$,
the $H_0$ purity increases with disorder
and depends weakly on $L$, as shown in
Fig.~\ref{fig:sdrg_purity}(b).
At $L=256$, it rises from $0.80$ at
$\delta=0.25$ to $0.96$ at $\delta=1$
and $0.98$ at $\delta=2$.
Nearest-neighbour purity remains at
$0.62$--$0.63$, since it misses the same
non-contiguous clusters at every $\delta$.
Meanwhile, non-nearest-neighbour merges
increase from $16\%$ to $43\%$.
Average linkage stays within $0.02$ of
single linkage, as do samples without
the fixed-mean shift.
At $\delta=3$ and $L\ge64$, the computed
correlations violate $0\le C_{ij}\le1$
by more than $10^{-3}$ because of numerical
precision loss in the free-fermion solution,
so these results are excluded.
The increasing agreement with SDRG as it
becomes exact indicates that the partial
recovery under box disorder reflects its
moderate strength rather than a limitation
of the correlation hierarchy.}

\begin{figure*}[t]
  \centering
  \includegraphics[width=0.95\textwidth]{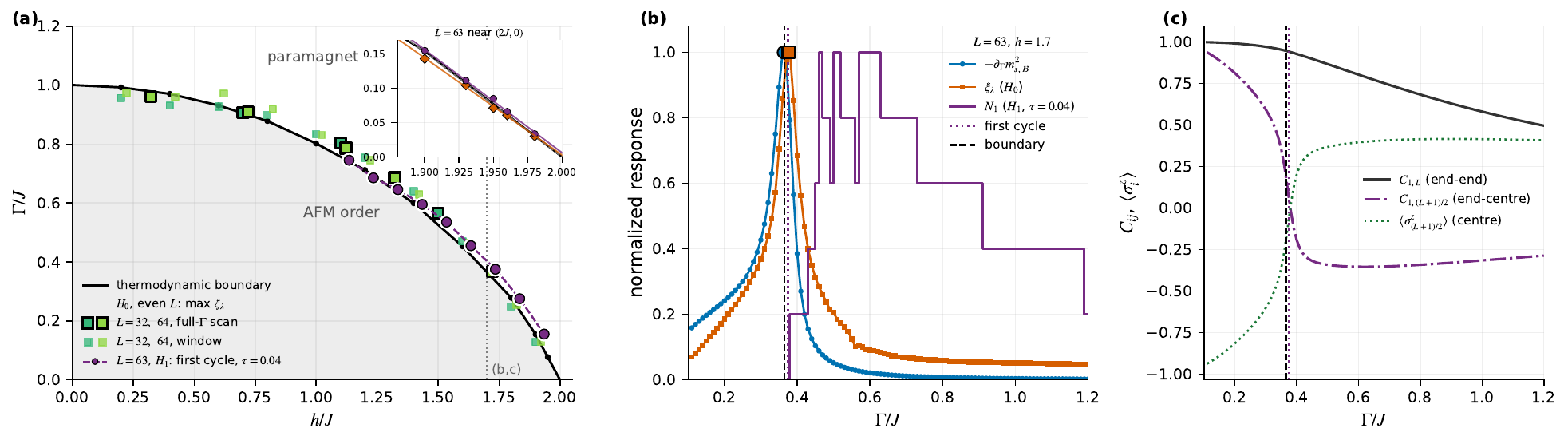}
\caption{\textbf{Clean chain in transverse and
longitudinal fields with open boundaries}.
(a) Phase diagram with the DMRG boundary of
Ref.~\cite{Lajko2020}, digitized from its Fig.~1.
Shading indicates the antiferromagnetic phase.
Squares show $\xi_\lambda$ maxima for even chains
$L=16$, $32$ and $64$, offset in $h$ for visibility.
Large thick-edged squares denote full scans over
$\Gamma/J\in[0,1.2]$, small translucent squares
restricted scans near the boundary, and open squares
maxima at window edges. Circles mark the onset of
$H_1$ cycles with lifetime above $\tau=0.04$
for odd chains $L=63$ (filled) and $31$ (open).
The inset shows the multicritical region at $L=63$,
with $\xi_\lambda$ maxima (diamonds), $H_1$ onsets
(circles), and the linear boundary
$\Gamma=1.526\,(2J-h)$ (dashed).
Solid lines fit Eq.~\eqref{eq:linear_fit}
for $h\ge1.9$.
(b) Full scan at $L=63$, $h=1.7$, indicated
by the dotted line in (a). Physical response
$-\partial_\Gamma m^2_{s,\mathcal B}$,
$\xi_\lambda$ and $N_1$, each normalized
to its maximum. Dotted and dashed lines mark
the first $H_1$ onset and reference boundary.
(c) End-to-end and end-to-centre correlations
$C_{1,L}$ and $C_{1,(L+1)/2}$, and central
magnetization $\langle\sigma^z_{(L+1)/2}\rangle$
along the same cut. Points with branch disagreement
or saturated bond dimension remain plotted but
are excluded from peak searches.}
  \label{fig:clean_frontier}
\end{figure*}

\rev{The advantage of $H_0$ originates in late
SDRG merges. We label each pair within a cluster
by the fraction $\rho$ of clusters still active
when SDRG joins its sites, from $\rho\simeq1$
for early merges to $\rho\to0$ for late ones,
and resolve purity by $\rho$.
For box disorder at $L=512$, early pairs
($\rho>1/4$) account for $7\%$ of all pairs
and have purity $0.96$ in both hierarchies.
Late pairs ($\rho\le1/32$) account for $70\%$
and have purity $0.75$ for $H_0$, against
$0.43$ for nearest-neighbour.
The gain therefore comes from late merges
across previously decimated sites.
At fixed $1/32<\rho\le1/16$, $H_0$ purity
increases from $0.67$ at $L=32$ to
$0.80$ at $L=512$, but longer chains
include later, less pure stages, explaining
the saturation of global purity.
For $\delta=2$, purity reaches
$0.975$--$0.986$ at every stage, showing
that stronger disorder improves the recovery
of late merges. Late-pair purity also includes
errors from earlier merges within each half,
rather than measuring late-stage accuracy alone.

These results use clusters from the complete
SDRG flow. Intermediate partitions are mostly
contiguous and are better recovered by the
nearest-neighbour hierarchy (Sec.~SII\,A of the Supplemental Material).
The trees themselves show a similar distinction.
At $L=512$, the probability that a same-cluster
pair has a lower cophenetic height than a
different-cluster pair is $0.96$ for $H_0$,
$0.85$ for nearest-neighbour and $0.70$
for unrelated clusters, rising to $0.997$
for $H_0$ at $\delta=2$.
Conversely, the nearest-neighbour tree better
reproduces the order of SDRG merges within a
cluster, since SDRG decimates nearest-neighbour
bonds of the renormalized chain.
Within clusters, correlations factorize into
site moments, whose ordering determines
the $H_0$ merges (Sec.~SII\,B of the Supplemental Material).}
This non-local structure identifies which sites
form strongly correlated clusters in individual
realizations, information absent from pseudocritical
points or pair-averaged correlations.

\section{Clean chain at finite longitudinal field}
\label{sec:clean_frontier}
% Fig. 5 (paper_figures/fig_clean_frontier). Numbers from paper_figures/fig_clean_frontier/output/fig_clean_frontier_summary.json.
% Supplement Figs. S3 (first H1 cycle) and S4 (robustness to tau) in the same folder.
% H1 rule and blind test: documentation/PLAN_CLEAN_FRONTIER_ODDL_CLOSURE_20261015.md (amendment v2).

At finite longitudinal field, the clean-chain transition
forms a mixed-order line in the $(h,\Gamma)$ plane,
connecting $(0,J)$ to $(2J,0)$, with a diverging
correlation length governed by Ising exponents and
discontinuous bulk correlations
~\cite{Ovchinnikov2003,LajkoIgloi2021}.
Near $(2J,0)$, the boundary is linear in
$2J-h$~\cite{Lajko2020}. Without an exact solution,
we use the DMRG boundary of Ref.~\cite{Lajko2020}
\rev{, digitized from its Fig.~1 with uncertainty
of about $0.006$ (Appendix~\ref{app:numerics})}.

We scan $\Gamma$ for even chains ($L=16$, $32$, $64$)
at $h=0.2$--$1.9$, mostly near the reference boundary.
For $L=64$, cuts at $h=0.3$, $0.7$, $1.1$, $1.3$
and $1.7$ cover $\Gamma/J\in[0,1.2]$.
We exclude $h\ge2J(1-1/L)$, where even open
chains are already ferromagnetic at $\Gamma=0$.
To avoid the boundary domain wall present for
even chains at $h>J$ (Sec.~\ref{sec:model}),
we also scan odd chains ($L=31$, $63$) over
the full range for $h=1.1$--$1.9$ in steps
of $0.1$. Maxima are located only at unflagged
points with unflagged neighbours.

\paragraph*{Response in $H_0$.}
Figure~\ref{fig:clean_frontier}(a) shows that the
$\xi_\lambda$ maxima for even chains follow the
reference boundary up to $h=1.9$, with mean
deviations decreasing from $0.055$ ($L=16$,
$9$ cuts) to $0.029$ ($L=32$, $13$ cuts)
and $0.023$ ($L=64$, $15$ cuts).
Since most scans target the reference boundary,
the full-range cuts provide an unguided test,
giving a deviation of $0.017$.
The physical response
$-\partial_\Gamma m^2_{s,\mathcal B}$ instead
deviates by $0.166$, $0.122$ and $0.091$,
and by $0.189$ in full-range scans, where
its maxima at $h=0.7$, $1.1$ and $1.3$
lie more than $0.1$ inside the ordered phase.
Odd chains reveal the role of the even-chain
boundary wall for $h>J$. Without it, the
physical response deviates by $0.015$ ($L=31$)
and $0.009$ ($L=63$), compared with $0.036$
and $0.020$ for $\xi_\lambda$.
Thus, $\xi_\lambda$ is more robust to parity,
although its advantage in even chains concerns
robustness rather than accuracy.
\rev{At $L=63$, most estimators deviate by
$0.008$--$0.013$ (Table~\ref{tab:benchmark}),
comparable to the reference uncertainty
and insufficient to rank them.}

Odd chains also reach the multicritical point,
since the N\'eel and polarized states are
degenerate at $h=2J$. For $L=63$, cuts at
$h=1.93$, $1.95$, $1.96$ and $1.98$,
with field steps down to $0.0025$, show
$\xi_\lambda$ maxima following the linear
boundary to $\Gamma=0.030$, as shown in
the inset of Fig.~\ref{fig:clean_frontier}(a).
Unweighted fits of the five cuts with
$h\ge1.9$ to
\begin{equation}
  \Gamma^\star(h)=a\,(h_c-h),
  \label{eq:linear_fit}
\end{equation}
give $h_c=2.002(2)\,J$, $a=1.41(4)$
for $\xi_\lambda$, and
$h_c=2.003(1)\,J$, $a=1.38(3)$
for the physical response.
Both extrapolate close to $h_c=2J$,
with slopes similarly below the hard-dimer
value $1.526$~\cite{Lajko2020}, suggesting
finite-size rather than observable-dependent
corrections.

Figure~\ref{fig:clean_frontier}(b) shows a
full-range cut at $L=63$, $h=1.7$, where
both responses peak at the boundary.
The landscape velocity remains finite
in the disordered phase as the correlation
hierarchy continues to evolve.
Its maximum therefore locates the
mixed-order transition, as at the
continuous transition for $h=0$, by detecting
rapid changes in correlations, including
their discontinuity.

The benchmark of Sec.~\ref{sec:peaks}, reported
in Table~\ref{tab:benchmark} of
Appendix~\ref{app:benchmark}, shows that
the merge tree departs from the lattice
near the boundary. Of its edges, $36\%$
($L=64$) and $57\%$ ($L=63$) cannot be
replaced by nearest-neighbour paths at
the same scale.
\rev{This non-locality arises from the
sublattice structure of the raw correlator.
At large distances,
$C_{ij}\simeq m_s^2+(-1)^{|i-j|}m_u^2$,
with $m_s$ and $m_u$ the staggered and
uniform magnetizations, plus a cross term
$\pm2m_sm_u$ at even distances.
Sites on the field-aligned sublattice can
therefore be more correlated than their
intervening nearest neighbours.
All non-replaceable edges connect sites
of this sublattice, with $91\%$ ($L=64$)
and $75\%$ ($L=63$) at distance two.
For connected correlations, these fractions
of non-replaceable edges fall to $5\%$
and $9\%.$}

This non-locality does not improve boundary
localization. The nearest-neighbour-bar
velocity and the derivatives of half-chain
entanglement entropy and nearest-neighbour
correlations are \rev{more accurate, with
deviations of $0.011$--$0.019$ versus $0.023$
for $\xi_\lambda$ at $L=64$, and
$0.008$--$0.013$ versus $0.020$ at $L=63$.}
The robustness of $\xi_\lambda$ in even chains
is relative to bulk magnetization and fidelity
susceptibility, both affected by the boundary
wall\rev{, while local responses are similarly
robust}.
The correlator choice also matters.
Using connected staggered correlations,
$\xi_\lambda$ deviates by only $0.008$
at $L=63$, but by $0.087$ at $L=64$,
where it inherits the bulk-magnetization
failure (Appendix~\ref{app:numerics} and
Table~\ref{tab:benchmark} of
Appendix~\ref{app:benchmark}).
Near the multicritical point, the energy
second derivative fails for both parities.
Low-cost spin flips of the classical state
produce maximal energy curvature at small
$\Gamma$, placing its peak at the lower
scan edge. This causes larger deviations
for $L=63$, scanned from $\Gamma=0$,
than for $L=64$, whose windows start
closer to the boundary.

\paragraph*{Response in $H_1$.}
Odd chains exhibit an additional $H_1$ signal.
More than $0.05$ inside the ordered phase, no
$H_1$ interval exceeds a lifetime of $0.014$.
Near the boundary, a persistent cycle emerges
and $N_1(\tau)$ increases
(Fig.~\ref{fig:clean_frontier}(b)).
We define its first onset as the $H_1$
pseudocritical point.

Figure~\ref{fig:clean_frontier}(c) reveals the
origin of this cycle. For widely separated sites,
Eq.~\eqref{eq:cstag} \rev{becomes}
$C_{ij}\simeq(-1)^{|i-j|}m_im_j$, with
$m_i=\langle\sigma_i^z\rangle$, accurate within
$7\times10^{-3}$ across all odd cuts.
The field aligns both ends, keeping
$C_{1,L}\simeq m_1^2$ large.
For $L=31$ and $63$, the centre
$c=(L+1)/2$ belongs to the opposite sublattice,
so $C_{1,c}\simeq-m_1m_c$ changes sign as
staggered order melts.
In the \rev{dissimilarity} $D$, the end
sublattice remains compact while the other
forms an arc that moves away at its centre,
enclosing a loop
(Fig.~\FigSuppHoneCycle{} of the Supplemental Material).
\rev{The magnetization profile alone cannot
produce this loop. Replacing correlations by
$(-1)^{|i-j|}m_im_j$, with unit diagonal,
eliminates all $H_1$ intervals across the
$22$ odd cuts of Fig.~\ref{fig:clean_frontier},
whereas full correlations reproduce every onset.
The product approximation differs from $C$
by less than $0.03$ beyond distance $L/2$,
but by $0.34$--$0.50$ for nearest neighbours.
Short-range connected correlations are therefore
also needed to form the sublattice arcs.}
The cycle thus encodes local order through the
global topology of $D$, explaining its accuracy
relative to bulk magnetization.
In even chains, the ends belong to opposite
sublattices and the loop cannot close.
No $H_1$ intervals occur in the wall-free
even scans at $L=64$, $h=0.3$ and $0.7$. \rev{In the odd chain with $L=31$ at $h=0$ and $0.5$} only
$H_0$ locates the transition \rev{(Sec.~SIII of the Supplemental Material)}.
The cycle is therefore specific to open odd
chains in a longitudinal field, rather than
a bulk observable.

We fixed $\tau=0.04$ and the onset rule
using $6$ odd cuts before analyzing the
remaining $12$, shown as circles in
Fig.~\ref{fig:clean_frontier}(a).
For $L=63$, all $7$ blind cuts locate
the boundary, with maximum and mean
deviations of $0.027$ and $0.008$,
the latter comparable to the physical response.
For $L=31$, $3$ of $5$ blind cuts agree,
while the others overestimate the boundary
by up to $0.079$.
Near the multicritical point, fitting
Eq.~\eqref{eq:linear_fit} gives
$h_c=2.004(2)\,J$ and $a=1.50(4)$,
consistent with the boundary slope.
The result is robust to $\tau$.
At $L=63$, the mean deviation remains
within twice its value at $\tau=0.04$
for $\tau=0.025$--$0.075$.
Onsets are missed only for $\tau\ge0.125$,
above the lifetime of the first cycle
(Fig.~\FigSuppHoneTau{}).

\section{Reentrant regime of the random chain}
\label{sec:reentrant}
% Fig. 7 (paper_figures/fig_reentrant, make_fig_reentrant_main.py). Numbers from fig_reentrant_main_summary.json and the analyses
% reentrant-oddL-L17-L33-v1/summary.json (odd L) and reentrant-exit-anatomy-v1/review_checks.json (even L).
% Story: documentation/HISTORIA_SECCION_REENTRANTE_20261014.md, corrected by
% RESPUESTA_REVISION_SECCION_REENTRANTE_20261014.md and RESULTADOS_REENTRANT_ODDL_20261016.md.
% Rings L16/L17 (\Pbc* macros): pbc-reentrant-ED-parity-v1/analysis/summary.json,
% documentation/RESULTADOS_PBC_ED_PARITY_MAIN_20261024.md.
% v3 D3 (2026-11-03): section shortened, details moved to the Supplemental Material, panel (c) added
% (prepare_parity_average.py). Previous version: git history of this file.

Unlike the monotonic boundary of the clean chain
(Sec.~\ref{sec:clean_frontier}), disorder produces
a reentrant transition near the random classical
multicritical point. Lajk\'o \textit{et al.}
show numerically and argue perturbatively that,
for $h$ slightly above
\rev{$2J_{\min}=2$, twice the minimum coupling
of the box distribution,} increasing $\Gamma_0$
first induces order in the disordered chain
and then destroys it~\cite{Lajko2020}.
This order-through-disorder mechanism provides
two boundaries with which to test the
landscape velocity.

\begin{figure}[!tbp]
  \centering
  \includegraphics[width=\columnwidth]{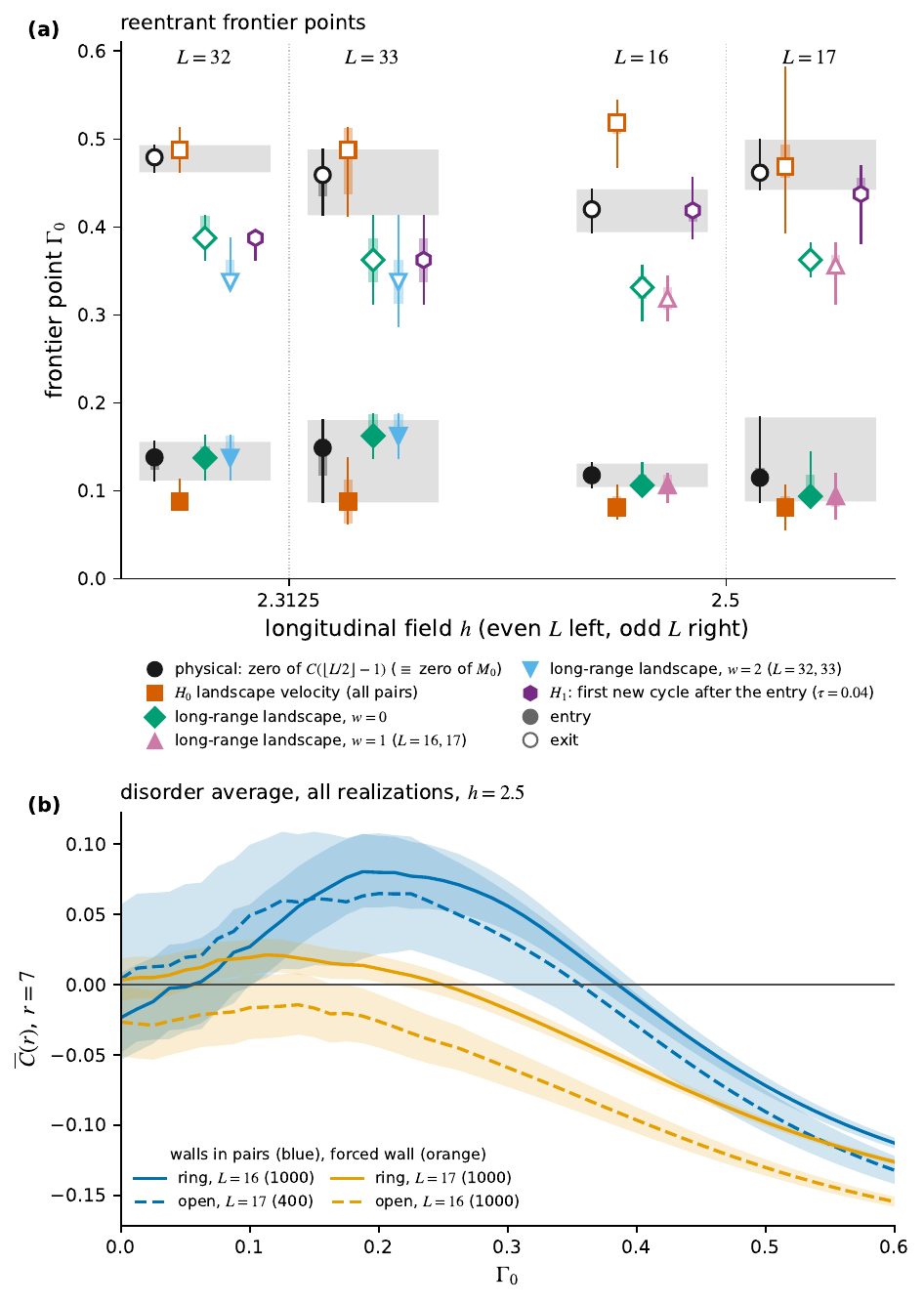}
\caption{\textbf{Reentrant regime of the random chain}.
(a) Reentrant realizations with open boundaries.
Even chains $L=16$ ($h=2.5$, $97/1000$ realizations)
and $L=32$ ($h=2.3125$, $137/1000$), and odd
wall-free chains $L=17$ and $33$ at the same fields
($40/400$ and $43/400$), plotted to the right
of the corresponding even sizes. Symbols show
median frontier estimates, with 95\% (whiskers)
and 50\% (thick bars) bootstrap intervals.
Grey bands mark the 95\% intervals of the physical
entry and exit, defined by zeros of $C(r)$ at
$r=\lfloor L/2\rfloor-1$.
Filled and open symbols denote entry and exit.
Estimators are the global $H_0$ velocity
(first and second maxima, the latter only when
interior), the long-range landscape velocity
with $w=0$ and $w=1$ ($L=16$, $17$) or
$w=2$ ($L=32$, $33$), using maxima before
and after the level apex, and the first
post-entry increase of $N_1$ for $\tau=0.04$,
as in Fig.~\ref{fig:clean_frontier}.
(b) Disorder-averaged $C(r)$ at $r=7$ and
$h=2.5$ for rings (exact diagonalization,
$1000$ realizations per size) and open chains
(DMRG, $1000$ for $L=16$ and $400$ for $L=17$),
with 95\% bootstrap bands. Blue denotes
walls in pairs and orange one forced wall.}
  \label{fig:reentrant}
\end{figure}

\paragraph*{\rev{Wall parity of the averaged lobe.}}
Lajk\'o \textit{et al.} define the reentrant boundary
by the zero of the disorder-averaged correlator
$\overline{C}(r)$ at $r=L/2-1$ in periodic
chains~\cite{Lajko2020}. Exact diagonalization
shows that this average depends on the parity
of classical domain walls at $\Gamma_0=0$
\rev{(Fig.~\ref{fig:reentrant}(b))}.
Even rings have an even number of walls and
reproduce the reentrant lobe~\cite{Lajko2020}.
In contrast, even open chains and odd rings
have a forced wall, suppressing the lobe,
with the average remaining negative in even
open chains.
\rev{In longitudinal-field scans, the zero-crossing
field $h^\ast$ shifts by $+0.118$
$[+0.022,+0.150]$ between $\Gamma_0=0$
and $0.225$ for $L=16$, confirming the
extension of the ordered region to larger $h$.
For $L=17$, the shift is only $+0.007$
$[-0.025,+0.039]$, consistent with zero  (Sec.~SIV\,A of the Supplemental Material).}
Thus, a forced wall suppresses the averaged
lobe regardless of boundary conditions.
Odd open chains instead have walls in pairs,
and the average starts near zero and changes
sign only once, yielding an exit but no entry.

\paragraph*{Realization-resolved reentrance.}
Although the disorder average is never reentrant
in our open chains ($L=16$--$33$), individual
realizations can be. We classify a realization as
reentrant when $C(r)$, with
$r=\lfloor L/2\rfloor-1$, is negative at both
scan endpoints and positive in between, with
minimum lobe amplitude and width
(Appendix~\ref{app:numerics}).
Approximately one in ten to one in seven
realizations satisfy this fixed criterion in
both open chains and rings\rev{, including
$97/1000$ ($L=16$) and $137/1000$ ($L=32$)
for open even chains, $40/400$ ($L=17$)
and $43/400$ ($L=33$) for open odd chains,
and $139/1000$ ($L=16$) and $98/1000$
($L=17$) for rings}.
The two zeros of $C(r)$ define the entry
and exit of each realization.
These are \rev{finite-size correlator sign
changes, not the averaged boundary of}
Ref.~\cite{Lajko2020}\rev{ or independent
estimates of a thermodynamic transition.
All results concern this selected subpopulation}.
\rev{We use $h=2.5$ for $L=16$, $17$
and $h=2.3125$ for $L=32$, $33$, within
the reentrant windows that approach $h=2$
with increasing size~\cite{Lajko2020}
(Sec.~SIV\,B of the Supplemental Material).
Since $r$ is odd, the large-distance
correlator satisfies $C(r)\simeq m_s^2-m_u^2$,
the difference between squared staggered
and uniform magnetizations.}

\paragraph*{Entry.}
Figure~\ref{fig:reentrant}(a) shows the estimator
medians over reentrant realizations.
The global $H_0$ velocity locates the entry
too early, at $\Gamma_0\approx0.081$--$0.087$
against physical medians of $0.115$--$0.149$,
with realization-resolved errors increasing
from $0.027$ ($L=16$) to $0.068$ ($L=32$).
Its forest is dominated by short-range
correlations that evolve throughout the scan,
whereas the boundaries are defined by
half-chain correlations.

We therefore construct a long-range landscape
from the $H_0$ barcode of $D$ restricted to
pairs with $||i-j|-r|\le w$.
For $w=0$, these pairs form a forest whose
barcode level $M_0=1-\langle d_a\rangle$
equals $C(r)$ exactly.
\rev{Its velocity is then a weighted norm of
the field derivatives of correlations at the
reference distance $r$, so the restriction
uses information from the reference itself.}
This velocity locates the entry within the
physical band for all four sizes, with errors
of $0.014$, $0.017$, $0.014$ and $0.016$.
Shells $w=1,2$ perform similarly.
The improvement comes from scale selection
rather than topology, since
$|\mathrm{d}C(r)/\mathrm{d}\Gamma_0|$ gives
errors of $0.008$--$0.014$.
\rev{Fidelity susceptibility also peaks at
the entry without scale selection, with
errors of $0.011$--$0.029$
(Sec.~SIV\,D of the Supplemental Material).}

\rev{The entry is abrupt, a narrow avoided crossing between configurations of
domain walls, and its slope is limited by the field step (Sec.~SIV\,B of the Supplemental Material). Only $17$ of the $317$ entries lower the total
magnetization by two, and in $16$ of them a domain of odd length
reverses and the two walls that bound it annihilate, the process of the
perturbative picture of Ref.~\cite{Lajko2020}. Most entries keep the
magnetization and displace a wall rather than remove it, in odd and even
chains alike (Table~\TabSuppWalls{} of the Supplemental Material).}

\paragraph*{Exit.}
None of the examined rate or response estimators
\rev{, including fidelity susceptibility,}
locates the exit \rev{(open symbols in
Fig.~\ref{fig:reentrant}(a))}, nor improves
on the constant median exit realization by
realization \rev{(Sec.~SIV\,F of the Supplemental Material)}.
The exit is a smooth crossing where the
long-range correlator $C(r)\simeq m_s^2-m_u^2$
vanishes as uniform magnetization overtakes
staggered order. Unlike the entry, its position
depends on $r$, and the ground-state energy
slope shows no abrupt change
(Fig.~\FigSuppBoundaries{} of the Supplemental Material).
It is an effective, size-dependent
point~\cite{Lajko2020}, not a collective
reorganization expected to produce a response
peak. The long-range landscape level $M_0$
marks it by construction, since $M_0=C(r)$.

\rev{Odd chains and rings show the same behavior.
Without the boundary wall, the exit is at least
twenty times less steep than the entry.
In reentrant rings, the long-range velocity
locates the entry, but no topological estimator
outperforms the constant baseline at the exit.}
Thus, velocity-based topology detects the abrupt
domain-wall reorganization at the entry when
the relevant scale is selected, with the
precision of the corresponding correlator,
but not the smooth exit.

\section{Discussion and outlook}
\label{sec:discussion}

We studied the persistence-landscape velocity of the
$H_0$ barcode of $D=1-C$ as a threshold-free
pseudocritical estimator across four regimes of the
antiferromagnetic Ising chain, benchmarking it against
conventional ground-state responses. Its maxima
extrapolate to the exact clean critical point
(Sec.~\ref{sec:clean_h0}), show comparable
sample-to-sample scaling in the random chain
\rev{and track global responses within individual
realizations} (Sec.~\ref{sec:random_peaks}),
follow the mixed-order transition line
(Sec.~\ref{sec:clean_frontier}), and detect
reentrant entry when restricted to the relevant
correlation scale (Sec.~\ref{sec:reentrant}).

The main added value of topology lies in the
correlation hierarchy. The $H_0$ barcode
\rev{records the merge scales of} the single-linkage
tree, determined by the minimum spanning tree of $D$.
At $h=0$ in the clean chain, this tree follows the
lattice, reducing the velocity to a weighted
bond-energy response with logarithmic critical growth.
Disorder and longitudinal fields can introduce
non-nearest-neighbour links into the correlation-based
tree. In the clean chain at finite longitudinal field,
these links preferentially connect sites on the
field-aligned sublattice, one of the two alternating
sets of sites. They reflect the sublattice structure
of the raw correlator, but do not improve the accuracy
with which the landscape velocity locates the phase
boundary.
In the random chain, however, the
\rev{labelled merge tree} recovers many of the
non-contiguous correlated clusters identified by the
strong-disorder renormalization group (SDRG) as subtrees.
Its dendrogram purity, which measures how well these
clusters are grouped without mixing in other sites,
approximately levels off as $L$ increases, whereas the
purity of nearest-neighbour hierarchies continues to
decrease.
\rev{At the largest size, the full tree is purer
in every realization.}
Average linkage performs equally well, whereas
a single correlation threshold loses information.
Thus, the essential information comes from
all pairwise correlations organized across scales.
\rev{Single linkage provides a stable,
threshold-free hierarchy whose merge scales
also define the $H_0$ barcode.}
It identifies the cluster membership of individual
realizations without running SDRG, beyond what
pseudocritical points or pair-averaged
correlations reveal (Sec.~\ref{sec:random_clusters}).

First homology provides a complementary signal.
In odd open chains under a longitudinal field,
a long-lived $H_1$ cycle marks the transition
for $h>J$. Correlated polarized ends and a
reorienting central spin produce a loop in
the correlation \rev{dissimilarity}.
This encodes local order through global topology
but depends on the open-chain geometry rather
than representing a bulk observable
(Sec.~\ref{sec:clean_frontier}).

The limitations are equally instructive.
Velocity detects abrupt reorganizations, including
domain-wall motion at reentrant entry, but not
the smooth correlator sign change at exit, which
none of the tested response estimators locates.
When short- and long-range correlations evolve differently
with the field, the velocity must be restricted to the
correlations relevant to the transition being probed.
The resulting estimator, however, is no more accurate
than the response of the corresponding correlator.
In random chains, pseudocritical widths approach
infinite-disorder scaling slowly for all estimators,
while SDRG cluster recovery remains partial at
moderate disorder \rev{but improves to a purity
of $0.98$ at $\delta=2$, as SDRG approaches
its asymptotically exact regime}.
These conclusions concern finite chains up to
$L=512$ at $h=0$ and $L=64$ at $h>0$.

The correlation hierarchy offers a route to
studying rare regions, Griffiths phases,
\rev{SDRG failures at moderate disorder,}
and higher-dimensional random systems.
The construction extends to other correlation
measures, including mutual information, and
experimental data. Periodic boundaries would
remove the boundary-induced $H_1$ signal,
while larger rings could connect individual
reentrance with the averaged phase boundary.
Already at $L=16$, even rings exhibit an
averaged reentrant lobe, absent in odd rings
with a forced wall (Sec.~\ref{sec:reentrant}).

\begin{acknowledgments}
F.B. acknowledges support from the Departament de
Recerca i Universitats de la Generalitat de Catalunya
and Universitat Ramon Llull.
\rev{N.S. acknowledges support from the Generalitat
de Catalunya (AGAUR, 2025 SGR 01604).}
Part of the computations were performed on the
Turing cluster of the Faculty of Mathematics and
Computer Science, University of Barcelona.
\end{acknowledgments}

\section*{Data availability}
\rev{The code and parameters required to reproduce
the results of this article, together with the
data underlying Figs.~\ref{fig:clean_h0}
and~\ref{fig:random_h0}, are publicly available
~\cite{BelchiStatutoCode2026}.
Other simulation outputs can be regenerated
using the released code and are available
from the authors upon reasonable request.}

\appendix
\section{Numerical details}
\label{app:numerics}

\paragraph*{DMRG protocol.}
DMRG calculations use TeNPy~\cite{Hauschild2018} (v1.1.0), with
$20$ two-site and $4$ single-site sweeps per field point, a
density-matrix mixer, energy and entropy tolerances of $10^{-10}$
and $10^{-8}$, and a singular-value cutoff of $10^{-12}$.
Each scan runs in both field directions. Its first point starts from
the classical ground state of Eq.~\eqref{eq:hamiltonian}, both N\'eel
states, a polarized state, and either the opposite polarized state
(clean) or a fixed-seed random product state (disordered), retaining
the lowest energy. Subsequent points use the preceding state.
The backward result replaces the forward one only if its energy is
lower by more than $10^{-12}$. We flag branch disagreements for
energy differences above $10^{-8}$ or infidelity above $10^{-6}$,
and bond saturation when the selected state reaches $\chi$.
Stored states provide the magnetizations used in
Secs.~\ref{sec:clean_frontier} and~\ref{sec:reentrant}.

\paragraph*{Datasets.}
Table~\ref{tab:datasets} lists DMRG sizes, bond dimensions, field
grids, scans and flags. At $h>0$, a scan is one window at fixed
$h$, potentially with several windows per cut. Of $108277$ field
points, $440$ ($0.4\%$) have branch disagreements and $238$
($0.2\%$) are bond-saturated. Random-chain saturation is counted
from the larger bond dimension of the two branches, giving an
overestimate.

\begin{table*}[t]
  \caption{DMRG datasets. The field is $g=\Gamma/J$ (clean) or
  $\Gamma_0$ (random), and step denotes the finest spacing.
  Branch and saturation columns count flagged points. Where
  multiple $\chi$ values are listed, the larger is used
  to assess saturation.}
  \label{tab:datasets}
  \begin{ruledtabular}
  \begin{tabular}{llrrcrrrrr}
  Fig. & system & $L$ & $\chi$ & field range & step & scans & points & branch & saturated \\
  \hline
  \ref{fig:clean_frontier} & clean, $h>0$ & 16 & 64 & $0$--$1.13$ & 0.01 & 15 & 501 & 0 & 0 \\
  \ref{fig:clean_frontier} & clean, $h>0$ & 31 & 128 & $0$--$1.2$ & 0.01 & 11 & 1331 & 16 & 0 \\
  \ref{fig:clean_frontier} & clean, $h>0$ & 32 & 128/256 & $0$--$1.2$ & 0.004 & 17 & 669 & 53 & 20 \\
  \ref{fig:clean_frontier} & clean, $h>0$ & 63 & 128 & $0$--$1.2$ & 0.0025 & 13 & 1373 & 0 & 0 \\
  \ref{fig:clean_frontier} & clean, $h>0$ & 64 & 128 & $0$--$1.2$ & 0.004 & 31 & 803 & 223 & 50 \\
  \ref{fig:reentrant} & random, reentrant & 16 & 32 & $0$--$0.6$ & 0.0125 & 1000 & 49000 & 4 & 43 \\
  \ref{fig:reentrant} & random, reentrant & 17 & 32 & $0$--$0.6$ & 0.0125 & 400 & 19600 & 0 & 82 \\
  \ref{fig:reentrant} & random, reentrant & 32 & 64 & $0$--$0.6$ & 0.025 & 1000 & 25000 & 87 & 27 \\
  \ref{fig:reentrant} & random, reentrant & 33 & 64 & $0$--$0.6$ & 0.025 & 400 & 10000 & 57 & 16 \\
  \end{tabular}
  \end{ruledtabular}
\end{table*}

\paragraph*{Quality flags.}
Flags remain in the data but are excluded, together with adjacent
points, when locating clean-chain maxima at $h>0$
(Sec.~\ref{sec:clean_frontier}). We exclude the unconverged
$L=64$, $h=1.95$ cut even at $\chi=512$.
\rev{All $278$ flagged points in the $L=32$ and $64$ cuts of
Fig.~\ref{fig:clean_frontier} lie at least $0.11$ below the
reference boundary, within the nearly degenerate ordered phase.
None lies within $0.05$ of a $\xi_\lambda$ maximum, the closest
being eight grid points away. Without masking, six of $28$
even-chain maxima shift to isolated flagged points.}
\rev{For random reentrant scans (Sec.~\ref{sec:reentrant}),
flagged points remain in peak and zero searches, comprising
at most $0.7\%$ of each dataset. Exact diagonalization of all
$L=16$ and $17$ reentrant realizations agrees with DMRG within
$10^{-12}$ in energy and $10^{-8}$ in entry and exit
(Sec.~SIV\,E of the Supplemental Material).}
None of the $61$ cuts at $h>0$ is bond-saturated within $0.1$
of the boundary, where the bond dimension stays below
$0.95\chi$. Increasing $\chi$ from $128$ to $256$ for
$L=32$, $h=1.9$ shifts the $\xi_\lambda$ and physical-response
maxima by only $4\times10^{-4}$ and $2\times10^{-4}$.

\paragraph*{Exact solution at $h=0$.}
The Jordan--Wigner mapping gives free fermions
~\cite{LiebSchultzMattis1961}. Using Majorana operators,
Eq.~\eqref{eq:hamiltonian} becomes
$\hat H=\mathrm i\,a^{\mathsf T}Tb$, with
$T_{jj}=\Gamma_j$ and $T_{j+1,j}=-J_j$.
The singular value decomposition $T=USV^{\mathsf T}$ gives
$W=UV^{\mathsf T}$, and $C_{i,i+r}$ is the $r\times r$ minor
of $-W$ formed by $-W_{l+1,k}$ for $k,l=i,\ldots,i+r-1$.
Computing all minors by Gaussian elimination without pivoting
agrees with pivoted LU within $2\times10^{-14}$ over $5052$
minors, for $L\le512$ in both phases and at criticality.
Ground-state fidelity is
$|\det[(W_1+W_2)/2]|^{1/2}$, choosing fermion parity to
maximize overlap when the lowest mode is degenerate.
We use this solution for clean $L=16$--$512$ and random
$L\le256$ chains ($1000$ realizations for $L\le128$,
$500$ at $L=256$), with field step $0.005$.
It agrees with exact diagonalization to machine precision.
DMRG correlations differ by at most $5\times10^{-7}$ (clean)
and $2\times10^{-5}$ (random, including flagged points) for
$L\le128$\rev{, using validation scans of step $0.025$
($0.0125$ for clean $L=128$) omitted from
Table~\ref{tab:datasets}}.
SDRG (Sec.~\ref{sec:random_h0}) decimates the largest field
or bond on the same samples.
\rev{The cluster comparison in Sec.~\ref{sec:random_clusters}
uses $300$ realizations per size for $L=32$--$512$ at
$\Gamma_0=1$. At $\delta=2$, $L=256$, the correlations
approach the precision limit. Repeating the
free-fermion calculation alters the purity in three of
$300$ realizations and its mean by $0.001$, while all
other purities agree to machine precision.}

\paragraph*{Topology.}
We obtain $H_0$ from the minimum spanning tree of $D$ using
Kruskal's algorithm and $H_1$ with GUDHI~\cite{Maria2014}
(v3.13.0), on the symmetrized Vietoris--Rips filtration of
$\mathrm{clip}(D,0,2)$, through dimension two with
$\mathbb{Z}_2$ coefficients. Landscapes use a uniform $t$ grid
covering all finite deaths. Field derivatives use second-order
finite differences in Secs.~\ref{sec:clean_h0},
\ref{sec:random_h0} and~\ref{sec:clean_frontier}.
For reentrance (Sec.~\ref{sec:reentrant}), forward differences
assign velocities to field midpoints, like interpolated zeros
of $C(r)$, without averaging across an abrupt entry.
At $h=0$ we compute $\xi_\lambda$ directly from
Eq.~\eqref{eq:velocity_deaths}\rev{, without a $t$ grid}.
\rev{At merge-scale crossings, centred differences average
one-sided derivatives and may underestimate the adjacent
one-sided velocities~\cite{BelchiStatuto2026}.}

Using finite differences of the ordered merge scales instead
of landscape layers changes $\xi_\lambda$ by at most $6\%$
in clean $h=0$ DMRG \rev{checks}, and refined maxima by at most
$0.007$, less than the \rev{scan steps of $0.0125$--$0.025$}.
\rev{This combines $t$-grid and merge-crossing effects.
For clean correlation error $\epsilon=5\times10^{-7}$,
the derivative error bound $2\epsilon/\Delta g$ is of order
$10^{-4}$ of the weighted root-mean-square merge-scale
derivative at a peak, using Eq.~\eqref{eq:velocity_deaths}.
For random error $2\times10^{-5}$, it stays below
$10^{-2}$ for derivatives of comparable magnitude.}
Doubling the exact-data field step to $0.01$ shifts clean
maxima by at most $10^{-3}$ and random maxima by a median
at most $0.003$. Near-tied peaks switch in $7\%$ of
$L=16$ and $1.6\%$ of $L=256$ samples, shifting by
more than $0.025$. Distribution widths change by at most
$5\%$, and their fitted exponent by $0.003$.
\rev{Since clean peaks sharpen as $1/L$, the maxima in
Fig.~\ref{fig:clean_h0}(c) and Table~\ref{tab:benchmark}
are recomputed on local exact-solution grids of step
$10^{-4}$. Halving the step moves them by less than
$10^{-6}$, while the correction to the original
coarser-grid maximum reaches $6\times10^{-4}$ for
$L=512$. Random-chain peaks retain their original
grid because this error is small compared with the
width of their distributions.}
All confidence intervals over realizations use percentile
bootstrap with $2000$ resamplings.

\rev{\paragraph*{Reference boundary.}
The reference for Sec.~\ref{sec:clean_frontier} consists
of $13$ points digitized from Fig.~1 of
Ref.~\cite{Lajko2020}, linearly interpolated in $h$.
The exact limits at $h=0$ and the multicritical
hard-dimer slope are reproduced within $0.001$.
Leave-one-out polynomial fits of degrees four to six
give root-mean-square errors of $0.005$--$0.008$,
including fit error. Monotone cubic interpolation
differs from linear interpolation by at most $0.006$
at the scanned fields. We therefore assign an
uncertainty of about $0.006$, excluding the intrinsic
DMRG finite-size and numerical errors of
Ref.~\cite{Lajko2020}. This is below the mean
deviation of each estimator but comparable to the
differences among the best at $L=63$, precluding
a meaningful ranking.}

\paragraph*{Choice of correlator.}
The maxima of $\xi_\lambda$ remain near the transition
when replacing the staggered correlator of
Eq.~\eqref{eq:cstag} with other choices.
At $h=0$ staggered correlations are positive,
making $|C|$ identical, while at $h>0$ it preserves
all maxima for odd $L=31$ cuts. Using unstaggered
$\langle\sigma_i^z\sigma_j^z\rangle$, also connected
at $h=0$, shifts the clean $L=16$ maximum from
$0.934$ to $0.903$, but by only $7\times10^{-5}$
at $L=128$. For random $h=0$ chains ($300$
realizations per size), median shifts fall from $0.028$
at $L=16$ to $0.008$ at $L=64$, with distribution
widths changing by at most $5\%$.
At $h>0$, the mean boundary deviations for odd
$L=31$, $63$ are $0.036$, $0.020$ with staggered
correlations and $0.035$, $0.018$ with unstaggered
ones. Connected staggered correlations,
$(-1)^{|i-j|}(\langle\sigma_i^z\sigma_j^z\rangle-
\langle\sigma_i^z\rangle\langle\sigma_j^z\rangle)$,
using stored-state magnetizations, improve them to
$0.013$, $0.008$ by removing the smooth field-induced
magnetization. For even $L=64$, however, the connected
input worsens the deviation from $0.023$ to $0.087$
and follows the bulk-magnetization failures. The
subtracted product carries the boundary wall and
depends on symmetry breaking in the selected state
(Table~\ref{tab:benchmark}). We therefore retain the
raw staggered correlator, whose long-distance limit
is the squared order parameter and is insensitive
to this state choice.

\paragraph*{Reentrant realizations.}
At $r=\lfloor L/2\rfloor-1$, a realization is selected
if $C(r)$ has exactly two zeros, its averages over
the first and last two field points are each below
$-0.02$, its maximum exceeds $0.05$, the zeros
are separated by at least $0.10$ in $\Gamma_0$,
and at least two consecutive grid points are positive.
Entry and exit are linearly interpolated zeros.
The rule was fixed at $L=16$ and applied unchanged
to $L=32$, $17$ and $33$.
Varying the amplitude threshold over $0.03$--$0.10$,
width $0.05$--$0.20$, endpoint margin $0.01$--$0.04$
and positive run length $1$--$4$ points, individually
or with all thresholds tightened, selects $70$--$106$
realizations at $L=16$ and $88$--$148$ at $L=32$,
with Jaccard overlap at least $0.64$ with the original
set. The conclusions persist. Across variants,
the slope of $C(r)$ locates entry within $0.016$
but misses exit by at least $0.058$, improving on
a constant by no more than $0.023$.
The selected samples are not atypical. At $L=16$,
the coupling spread is $0.275$ versus $0.277$ overall,
with $3.0$ versus $2.7$ walls at $\Gamma_0=0$.
At $L=32$, wall counts are $3.2$ versus $3.1$.

\paragraph*{Use of AI.}
\rev{Parts of the analysis code were implemented
with assistance from AI coding tools
(OpenAI Codex and Anthropic Claude), following
author-defined procedures. The authors reviewed
all code and take full responsibility for the
reported results.}

\section{Benchmark of the pseudocritical estimators}
\label{app:benchmark}

Table~\ref{tab:benchmark} compares the estimators
of Sec.~\ref{sec:peaks} across the clean and random
chains at $h=0$ and the clean chain at $h>0$
(Secs.~\ref{sec:clean_h0}, \ref{sec:random_h0}
and~\ref{sec:clean_frontier}).

\begin{table*}[t]
  \caption{Benchmark of pseudocritical estimators
  using the same peak rule.
  For clean $h=0$, the peak position $g^\star$
  at $L=256$ (exact $g_c=1$)
  \rev{is refined off the field grid
  (Appendix~\ref{app:numerics})}.
  For random $h=0$, the distribution width
  $\sigma$ at $L=256$ ($500$ realizations)
  and its scaling exponent $p_\sigma$ from
  $L=16$--$256$ are shown with 95\% bootstrap
  intervals (IDFP value $p_\sigma=1/2$).
  For clean $h>0$, entries are mean absolute
  deviations from the reference boundary for
  even ($L=64$) and odd ($L=63$) chains.
  The second row uses connected staggered
  correlations, identical to the raw input
  at $h=0$ since $\langle\sigma_i^z\rangle=0$.}
  \label{tab:benchmark}
  \begin{ruledtabular}
  \begin{tabular}{lccccc}
  estimator & clean, $h=0$ & \multicolumn{2}{c}{random, $h=0$} & \multicolumn{2}{c}{clean, $h>0$} \\
   & $g^\star(256)$ & $\sigma(256)$ & $p_\sigma$ & $L=64$ & $L=63$ \\
  \hline
  $\xi_\lambda$ (merge scales) & 0.9945 & 0.0162 & 0.38 [0.35, 0.41] & 0.023 & 0.020 \\
  $\xi_\lambda$ (connected input) & 0.9945 & 0.0162 & 0.38 [0.35, 0.41] & 0.087 & 0.008 \\
  $\xi_\lambda$ (nearest-neighbour bars) & 0.9945 & 0.0152 & 0.25 [0.22, 0.28] & 0.019 & 0.013 \\
  $-\partial m^2_{s,\mathcal B}$ & 0.9943 & 0.0155 & 0.27 [0.24, 0.30] & 0.091 & 0.009 \\
  $|\partial\,\overline{C}_{i,i+1}|$ & 0.9937 & 0.0148 & 0.17 [0.14, 0.20] & 0.014 & 0.010 \\
  $\|\partial C\|_F/L$ & 0.9926 & 0.0158 & 0.16 [0.13, 0.20] & 0.093 & 0.012 \\
  $-\partial^2 e$ & 0.9935 & 0.0160 & 0.19 [0.15, 0.24] & 0.051 & 0.094 \\
  $|\partial S_{L/2}|$ & 0.9967 & 0.0256 & 0.18 [0.13, 0.22] & 0.011 & 0.008 \\
  $\chi_F$ & 0.9928 & 0.0152 & 0.18 [0.15, 0.22] & 0.061 & 0.012 \\
  \end{tabular}
  \end{ruledtabular}
\end{table*}

% \nocite{Fortunato2007}
\bibliography{references}

\ifarxiv
\clearpage
\begin{center}
  \textbf{\large Supplemental Material for ``Zeroth persistent homology of
  ground-state correlations at clean and disordered quantum transitions''}
\end{center}

\setcounter{section}{0}
\setcounter{subsection}{0}
\setcounter{equation}{0}
\setcounter{figure}{0}
\setcounter{table}{0}

% Que las ecuaciones no se reinicien con cada sección (lo impone el modo apéndice)
\counterwithout{equation}{section}

\renewcommand{\thesection}{S\Roman{section}}
\renewcommand{\thesubsection}{\thesection\,\Alph{subsection}}
\renewcommand{\theequation}{S\arabic{equation}}
\renewcommand{\thefigure}{S\arabic{figure}}
\renewcommand{\thetable}{S\arabic{table}}

% Anclas únicas para hyperref
\renewcommand{\theHsection}{S\Roman{section}}
\renewcommand{\theHsubsection}{S\Roman{section}.\Alph{subsection}}
\renewcommand{\theHequation}{S\arabic{equation}}
\renewcommand{\theHfigure}{S\arabic{figure}}
\renewcommand{\theHtable}{S\arabic{table}}

% Encabezados del SM sin el prefijo "Appendix"
\newcommand{\smsection}[1]{%
  \refstepcounter{section}\section*{\thesection.\quad #1}}
\newcommand{\smsubsection}[1]{%
  \refstepcounter{subsection}\subsection*{\Alph{subsection}.\quad #1}}

\rev{Reference numbers refer to the main text.}%

% v3 B9: scripts/analyze_unweighted_velocity_review_v3.py, results/review_v3/unweighted-velocity-v1/summary.json (\Uv* macros)
\rev{\smsection{Velocity without the landscape weights}
\label{sec:supp_v0}

Equation~\eqref{eq:velocity_deaths} weights the speed of every merge scale $d_{k}$
with the scale itself. Without this weight the velocity becomes
\begin{equation}
  v_0=\Bigl[\frac{1}{L-1}\sum_{k=1}^{L-1}\bigl(\partial_g d_{k}\bigr)^2\Bigr]^{1/2},
  \label{eq:supp_unweighted}
\end{equation}
the speed of the sorted edge lengths of the minimum spanning tree\rev{,
comparable to $\xi_\lambda/\sqrt{L-1}$ in Fig.~\ref{fig:clean_h0}}. Most merge
scales are short, so $v_0$ is dominated by the strongest correlations, which in
the clean chain at $h=0$ are those between nearest neighbours. We locate its
maxima with the same rules as those of $\xi_\lambda$. In the clean chain at
$h=0$ they approach $g_c=1$ with local exponents between $1.01$ and
$1.03$ for $L=16$ to $512$, compatible with $1/\nu=1$, and for $L=256$ the
maximum lies at $g^\star=0.9939$. The local exponents of $\xi_\lambda$, from
$0.88$ to $0.95$ (Sec.~\ref{sec:clean_h0}), differ from $1$
because of the weights. In the random chain at $h=0$ the maxima of $v_0$
follow those of $\xi_\lambda$ realization by realization, with Pearson
correlations between $0.78$ and $0.85$ for $L=32$ to $256$ and a median
difference of $0.001$. Their width for $L=256$ is $0.0165$ and decreases
with $p_\sigma=0.34$ $[0.31,0.37]$, compatible with that of
$\xi_\lambda$ (Table~\ref{tab:benchmark}). Their correlation with the maxima of
the physical response is lower, between $0.33$ and $0.44$ against
$0.44$ to $0.56$ for $\xi_\lambda$. On the clean boundary at
$h>0$ the mean deviation of the maxima from the reference is $0.017$
for $L=64$, $0.025$ for $L=31$ and $0.016$ for $L=63$,
against $0.023$, $0.036$ and $0.020$ for
$\xi_\lambda$, and close to the local responses of Table~\ref{tab:benchmark}.
In the short even chains it is larger, $0.120$ and $0.043$
for $L=16$ and $32$ against $0.055$ and $0.029$. Neither
version is more accurate in all regimes, so the weights of the landscape are
not needed to locate the transitions. They shift the velocity from the short
bars to the long ones, which record the weakest links of the correlation
hierarchy, and they make it the speed of the landscape in $L^2$
(Sec.~\ref{sec:landscapes}).}

\rev{\smsection{Clusters of the strong-disorder renormalization group}
\label{sec:supp_sdrg}

\smsubsection{Clusters of the stopped renormalization-group flow}

The clusters compared with the $H_0$ hierarchy in
Sec.~\ref{sec:random_clusters} are those of the complete flow of the
strong-disorder renormalization group. If the flow is stopped when a fraction
$\rho_{\rm stop}$ of the clusters is still active, and each active cluster is
counted as a cluster, the resulting partition refines the final one and its
pieces are mostly contiguous. The nearest-neighbour hierarchy recovers it
better. For $\rho_{\rm stop}=1/4$ at $L=512$ its purity is $0.91$
against $0.80$ for $H_0$, and for the power-law distribution with
$\delta=2$ at $L=256$ the
$H_0$ purity of the stopped partitions, for $\rho_{\rm stop}$ from $1/2$ to
$1/16$, falls to between $0.50$ and $0.60$, against
$0.86$ to $0.91$ for the nearest-neighbour hierarchy and
$0.98$ for the complete flow. The ground state at $\Gamma_0=1$ is
the end point of the flow and keeps little memory of the order in which each
cluster was assembled. Within a final cluster the median of $D$ between
different pieces of the partition stopped at $\rho_{\rm stop}=1/4$ is
$0.52$, against $0.32$ within one piece, for the box
distribution, and the contrast nearly disappears for $\delta=2$,
with $0.098$ against $0.086$. The $H_0$ hierarchy
therefore recovers the clusters formed by the complete flow, the sites
decimated together by one field, and not the partitions at intermediate
scales.}

\rev{\smsubsection{Comparison of the trees}

The renormalization group also orders the sites of each cluster, since every
bond decimation joins two of its parts at a lower energy scale than the
previous one. We compare this hierarchy with the trees built from $D$ through
their cophenetic heights, the scale at which a tree first joins two sites. For
the renormalization group the height of a pair of the same cluster is the stage
$\rho$ of the flow at which it is joined (Sec.~\ref{sec:random_clusters}), and
pairs of different clusters lie above all of them. Since the energy scale
decreases monotonically along the flow, the order of $\rho$ is that of
$\ln\Omega$, and we use rank measures that need no choice of scale. At the
level of clusters, the measure is the probability that a pair of the same
cluster has a lower height in the tree than a pair of different clusters (AUC). Within
a cluster, it is the Spearman correlation between the heights in the tree and
the order of the joins, averaged over the clusters with at least three sites
with weights equal to their number of pairs. Besides the $H_0$,
average-linkage and nearest-neighbour trees and the $H_0$ tree against the
clusters of another realization, we include as a reference the heights
$|i-j|$, which contain no correlations.

Table~\ref{tab:supp_trees} gives both measures for $300$ realizations per row.
At the level of clusters the $H_0$ and average-linkage trees agree best with
the renormalization group, and their advantage over the nearest-neighbour tree
grows with $L$ and $\delta$, as for the dendrogram purity of
Fig.~\ref{fig:sdrg_purity}. The order of the joins within a cluster is
followed by the nearest-neighbour tree and not by the $H_0$ tree. Part of this
order is set by the distance, since the renormalization group joins distant
sites late, but given $|i-j|$ the partial rank correlation with the order of the
joins is still between $0.73$ and $0.77$ for the
nearest-neighbour tree at $L=256$, against $0.34$ for $H_0$ in the box
distribution and $0.03$ for $\delta=2$. \rev{The distance also explains why
the $H_0$ tree follows the joins of another realization, with $0.40$ to
$0.49$, better than its own at $L=512$ and for $\delta\ge0.25$.} This is
expected. The renormalization group joins a cluster by decimating bonds of the
renormalized chain, which \rev{link clusters adjacent in that chain, separated
in the original chain only by decimated sites}, and the nearest-neighbour
correlations follow these bonds. The ground state, instead, does not keep the
order of the joins in its correlations. Within a cluster with at least four
sites, the fit $\ln C_{ij}=\mu_i+\mu_j$ explains a fraction $0.56$ of the
variance of $\ln C_{ij}$ for the box distribution, $0.90$ for
$\delta=1$ and $0.94$ for $\delta=2$ at $L=256$, so that the correlations
factorize into moments $e^{\mu_i}$ of the sites. The single-linkage height of
two sites is then set by the weaker of them, with a rank correlation of
$0.58$ with $-\min(\mu_i,\mu_j)$ for the box distribution and $0.96$
for $\delta=2$. The $H_0$ hierarchy thus records which sites form each cluster
and how strongly each site belongs to it, but not the order in which its bonds
were decimated.

\begin{table*}[t]
  \caption{Comparison of the trees of $D$ with the hierarchy of the
  strong-disorder renormalization group at $h=0$, $\Gamma_0=1$, $300$
  realizations per row. AUC is the probability that a pair of sites of the same
  final cluster has a lower cophenetic height in the tree than a pair of
  different clusters (ties count one half). ``Within'' is the Spearman
  correlation between the heights and the order in which the
  renormalization group joins the sites of a cluster, averaged over clusters
  with at least three sites with weights equal to their number of pairs.
  Trees: $H_0$ (single linkage), average linkage, nearest-neighbour links,
  the distance $|i-j|$, and the $H_0$ tree against the clusters of another
  realization (unrelated).}
  \label{tab:supp_trees}
  \begin{ruledtabular}
  \begin{tabular}{lrcccccccccc}
  & & \multicolumn{5}{c}{AUC} & \multicolumn{5}{c}{within} \\
  disorder & $L$ & $H_0$ & average & NN & $|i-j|$ & unrelated & $H_0$ & average & NN & $|i-j|$ & unrelated \\
  \hline
  box & 32 & 0.936 & 0.939 & 0.933 & 0.815 & 0.700 & 0.77 & 0.78 & 0.78 & 0.73 & 0.42 \\
  box & 64 & 0.931 & 0.938 & 0.916 & 0.814 & 0.704 & 0.77 & 0.78 & 0.83 & 0.77 & 0.48 \\
  box & 128 & 0.933 & 0.944 & 0.886 & 0.801 & 0.695 & 0.73 & 0.74 & 0.88 & 0.80 & 0.48 \\
  box & 256 & 0.944 & 0.950 & 0.864 & 0.801 & 0.693 & 0.59 & 0.64 & 0.91 & 0.82 & 0.49 \\
  box & 512 & 0.964 & 0.966 & 0.846 & 0.801 & 0.699 & 0.44 & 0.52 & 0.89 & 0.82 & 0.47 \\
  $\delta=0.25$ & 256 & 0.962 & 0.959 & 0.865 & 0.805 & 0.695 & 0.41 & 0.54 & 0.88 & 0.80 & 0.44 \\
  $\delta=0.5$ & 256 & 0.985 & 0.982 & 0.862 & 0.805 & 0.686 & 0.20 & 0.31 & 0.89 & 0.80 & 0.40 \\
  $\delta=0.75$ & 256 & 0.991 & 0.990 & 0.860 & 0.805 & 0.688 & 0.14 & 0.20 & 0.90 & 0.80 & 0.40 \\
  $\delta=1$ & 256 & 0.995 & 0.994 & 0.859 & 0.805 & 0.688 & 0.11 & 0.15 & 0.90 & 0.80 & 0.40 \\
  $\delta=1.5$ & 256 & 0.997 & 0.997 & 0.856 & 0.805 & 0.689 & 0.08 & 0.11 & 0.89 & 0.80 & 0.40 \\
  $\delta=2$ & 256 & 0.997 & 0.998 & 0.855 & 0.805 & 0.688 & 0.07 & 0.09 & 0.89 & 0.80 & 0.40 \\
  \end{tabular}
  \end{ruledtabular}
\end{table*}}

% v3: scripts/sdrg_modularity_cut_review_v3.py, results/review_v3/sdrg-modularity-cut-v1/summary.json (\Hm* macros)
\rev{\smsubsection{A single cut against a multiscale selection of the tree}

Hémery \textit{et al.} identify correlation clusters by removing the weakest
bonds of the correlation graph one by one and keeping the partition of
maximal modularity~\cite{Hemery2022}. Their sequence of partitions is the
single-linkage hierarchy read from the top, so their method is one cut of our
$H_0$ tree. We apply it to the same realizations as in the main text, with
weights $M_{ij}=C_{ij}$, the similarity that corresponds to $D=1-C$, and update
the modularity at every merge of the tree. As a selection without reference
at every scale we use HDBSCAN~\cite{Campello2013} in the implementation of
Ref.~\cite{Pedregosa2011}. With a minimal number of samples of one its
reachability distance is $D$ itself, so that it works on the same tree. With a
minimal cluster size of two and the excess-of-mass rule it keeps the subtrees
of largest stability, the summed persistence of their sites, at any scale, and
it leaves the other sites as single sites. Both parameters were fixed before
the comparison. Measuring the persistence in $-\ln C$ instead of $D$ changes
the scores by less than $10^{-3}$. Table~\ref{tab:supp_cuts} compares both
partitions with the clusters of the renormalization group, with the
size-weighted Jaccard index of the main text, and with the two bounds that use
the reference, the best single threshold and the best subtrees.

The cut of maximal modularity keeps few blocks, about ten at $L=512$ against
$257$ clusters, as expected from the resolution limit of the
modularity\rev{~\cite{Fortunato2007}}, and its score falls from $0.49$ at $L=32$ to $0.22$ at
$L=512$. The clusters of the renormalization group form at scales that differ
by orders of magnitude, so that no single level of the tree contains them.
HDBSCAN takes each subtree at the scale at which it is most stable. Its score
falls less, from $0.58$ to $0.43$, and the paired difference to
the modularity cut grows with $L$, to $0.21$
$[0.20,0.22]$ at $L=512$. Both remain well below the
$0.76$ of the best subtrees, so that a selection without reference
recovers only part of the clusters contained in the hierarchy.

\begin{table*}[t]
  \caption{Partitions of the $H_0$ tree against the clusters of the
  renormalization group, $h=0$, $\Gamma_0=1$, box distribution, $300$
  realizations per row. Size-weighted Jaccard index of the cut of maximal
  modularity (Mod.) and of HDBSCAN (HDB), their paired difference, the number
  of realizations in which HDBSCAN is higher (wins), the mean numbers of
  blocks of the cut and of clusters of the renormalization group, and the two
  bounds that use the reference (best single threshold and best subtrees).
  \rev{The clusters include single sites. As in the main text, the Jaccard
  index is computed over the clusters with at least two sites.}}
  \label{tab:supp_cuts}
  \begin{ruledtabular}
  \begin{tabular}{rcccrrrcc}
  $L$ & Mod. & HDB & HDB$-$Mod. & wins & blocks & clusters & threshold & subtrees \\
  \hline
  32 & 0.49 & 0.58 & $+0.09$ & 196 & 4.7 & 17 & 0.76 & 0.83 \\
  64 & 0.38 & 0.56 & $+0.17$ & 248 & 5.4 & 34 & 0.71 & 0.80 \\
  128 & 0.33 & 0.50 & $+0.17$ & 256 & 7.1 & 66 & 0.67 & 0.77 \\
  256 & 0.27 & 0.45 & $+0.18$ & 262 & 9.8 & 130 & 0.65 & 0.76 \\
  512 & 0.22 & 0.43 & $+0.21$ & 289 & 10.5 & 257 & 0.64 & 0.76 \\
  \end{tabular}
  \end{ruledtabular}
\end{table*}}

\smsection{The first \texorpdfstring{$H_1$}{H1} cycle of the clean odd chains}
\label{sec:supp_h1}

Figure~\ref{fig:supp_h1_cycle} draws the cycle of Sec.~\ref{sec:clean_frontier}
in one cut, $L=31$ at $h=1.7\,J$, where the onset lies at
$\Gamma=0.445\,J$ and the reference boundary at $0.366\,J$. Before the
onset the longest $H_1$ interval lives at most $0.032$. The sites are
placed by classical multidimensional scaling of $D$, which only serves to draw
them, while the persistence is computed on $D$ itself. The sublattice of the
ends, polarized along the field, forms a compact cluster that contains both
ends, $D_{1,L}=0.13$ in panel (b). The opposite sublattice, to which the
central site belongs, forms an arc whose middle lies far from the ends,
$D_{1,(L+1)/2}=1.30$. \rev{Figure~\ref{fig:supp_h1_cycle}(a) shows that
before} the onset the sites of the arc are linked
to the cluster by many short edges and the loop is filled. After it these
edges lengthen, and a cycle along the arc that closes through the cluster\rev{,
drawn in Fig.~\ref{fig:supp_h1_cycle}(b),} is born at $D=0.736$ and
dies at $0.802$. It is born when the
edge between sites 25--26 attaches one side of the arc to the
cluster, and it dies when the triangle of sites 7, 8 and 9 fills the
other side. The cycle drawn is this birth edge closed by the shortest path
between its two sites at the scale of birth.

At the onset the end spins are polarized with $\langle\sigma^z_1\rangle\ge
0.84$ in all odd cuts, and the correlation between the two ends is at
least $0.72$ for $L=63$. The onset precedes the change of
sign of $\langle\sigma^z_{(L+1)/2}\rangle$ by at most $0.075$ for
$L=63$. The deviation of the onset from the boundary decreases from
$L=31$ to $63$ in $6$ of the $9$ fields, a trend
rather than an extrapolation. Two control scans with $L=31$ show that the
cycle needs ends polarized by the field. At $h=0$ no $H_1$ interval appears in
the whole scan. At $h=0.5$, where $C_{1,L}$ falls to $0.10$, only a
short cycle appears (lifetime at most $0.058$), at
$\Gamma=1.055$, well above the boundary at $0.951$.

Figure~\ref{fig:supp_h1_tau} repeats the onset rule on the nine cuts
$h=1.1$ to $1.9$ for thresholds $\tau$ from $0.005$ to $0.15$. For
$L=63$ the mean deviation from the boundary is $0.009$ at
$\tau=0.04$ and stays within twice this value for $\tau$ between
$0.025$ and $0.075$. For $L=31$ it is
$0.034$, within twice this value for $\tau$ between
$0.005$ and $0.060$. For larger thresholds the onset
moves into the disordered phase, and some cuts have no onset from
$\tau=0.075$ ($L=31$) and $0.125$ ($L=63$) on,
when $\tau$ exceeds the lifetime of the first cycle. The range of
acceptable thresholds shifts to larger $\tau$ with the size, as the
lifetime of the cycle grows.

\begin{figure*}[t]
  \centering
  \includegraphics[width=\textwidth]{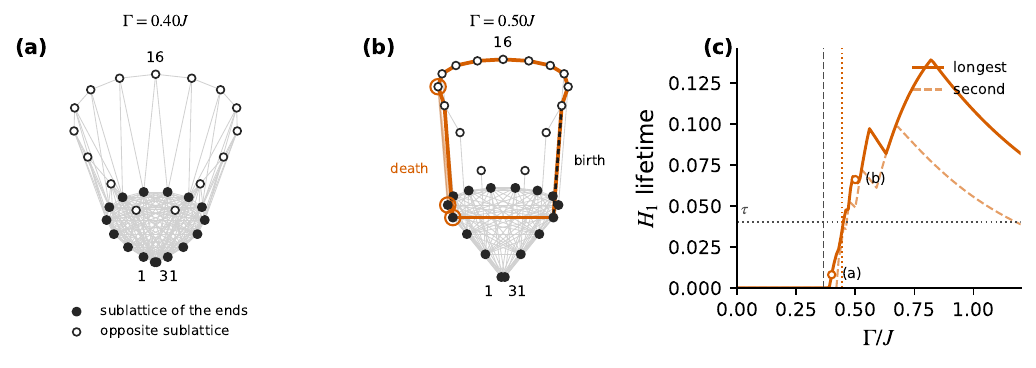}
  \caption{The first $H_1$ cycle in the clean odd chain, $L=31$ and
  $h=1.7\,J$. (a),~(b)~Sites placed by classical multidimensional scaling
  of $D=1-C$ at $\Gamma=0.40\,J$, before the onset, and $\Gamma=0.50\,J$,
  after it. Filled circles mark the sublattice of the ends (sites $1$ and
  $31$), open circles the opposite sublattice, which contains the central
  site. Grey lines are the edges with $D_{ij}$ below the midpoint of the
  lifetime of the cycle in (b), the same scale in both panels. The thick line
  is the representative of the longest $H_1$ interval, the dashed segment the
  edge at which it is born and the shaded triangle, with ringed vertices, the
  triangle at which it dies. (c)~Lifetimes of the
  two longest $H_1$ intervals against $\Gamma$. The horizontal dotted line is
  the threshold $\tau=0.04$, the vertical dotted line the onset and the
  dashed line the reference boundary. Circles mark the points of (a) and (b).}
  \label{fig:supp_h1_cycle}
\end{figure*}

\begin{figure}[t]
  \centering
  \includegraphics[width=\columnwidth]{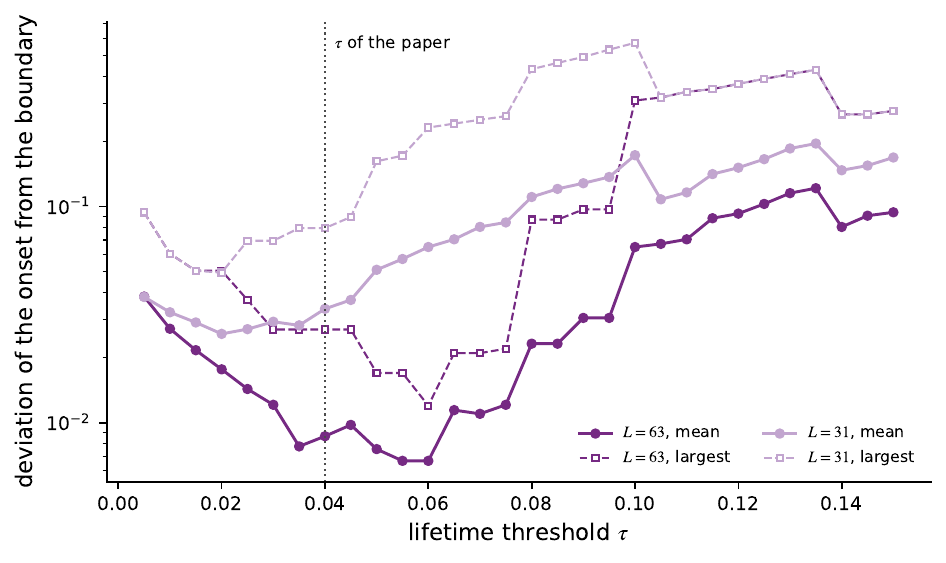}
  \caption{Onset of the first $H_1$ cycle against the lifetime threshold
  $\tau$, for the odd chains $L=31$ and $63$ on the cuts $h=1.1$ to $1.9$.
  Mean (circles) and largest (squares) absolute deviation from the reference
  boundary. The dotted line marks the threshold used in
  Sec.~\ref{sec:clean_frontier}.}
  \label{fig:supp_h1_tau}
\end{figure}

\smsection{The reentrant regime}
\label{sec:supp_reentrant}

% source: scripts/analyze_pbc_reentrant_ED_parity.py (M_h_map), results/random_ising_gamma_ensemble/pbc-reentrant-ED-parity-v1/analysis/summary.json
\smsubsection{Field scans of the rings}

The values of $h^\ast$ in Sec.~\ref{sec:reentrant} come from exact
diagonalization of rings with $L=16$ and $17$, for $200$ realizations per size,
independent of those of Fig.~\ref{fig:reentrant}(b), at $h=2.40$, $2.45$,
$2.50$, $2.55$ and $2.60$ with the same disorder at every $h$. At each $h$ the
Lanczos method, with tolerance $10^{-10}$ and started from the ground state of
the previous point, scans $\Gamma_0$ from $0$ to $0.6$ in steps of $0.0125$.
For each $\Gamma_0$ we average $C(r)$, $r=7$, over the realizations, make the
average non-increasing in $h$ by isotonic regression, and take $h^\ast$ as its
zero by linear interpolation between the two fields that bracket it. The value
$\Gamma_0=0.225$ was fixed before the scan from earlier simulations of the
rings. The interval is the 95\% percentile interval of
$h^\ast(0.225)-h^\ast(0)$ over $1000$ bootstrap resamples of the
realizations. For $L=16$ the zero at $\Gamma_0=0$, $h^\ast=2.428$, lies close
to the lowest field, and the $16\%$ of resamples in which it falls below
$h=2.40$ are discarded, which shortens the upper side of the interval. For
$L=17$ this happens in $0.5\%$ of the resamples.

\rev{\smsubsection{The two boundaries of the reentrant lobe}}

Figure~\ref{fig:supp_boundaries} collects the physical observables behind the
statements of Sec.~\ref{sec:reentrant} on the entry and the exit of individual
reentrant realizations. It uses the same selected realizations as
Fig.~\ref{fig:reentrant}, the even chains $L=16$ and $L=32$ and the odd chains
$L=17$ and $L=33$, at $h=2.5$ for $L=16,17$ and $h=2.3125$ for
$L=32,33$, with $r=\lfloor L/2\rfloor-1$. The local magnetizations were
measured on the saved ground states, without new DMRG runs. Numbers are given
in the order $L=16$, $17$, $32$, $33$. \rev{The fields were fixed for each even
size in exploratory scans, before the production runs. They depend on the size
because the finite-size transition points of Ref.~\cite{Lajko2020} move
towards $h=2$ as $L$ grows, as shown in its Fig.~5, and, as in that
reference, each field is placed inside the reentrant window of its own size.
The distance $r$ is odd for all four sizes, so the cross term $\pm2m_sm_u$ of
even distances cancels and $C(r)\simeq m_s^2-m_u^2$ at large distance.}

\rev{Figure~\ref{fig:supp_boundaries}(a) aligns the correlator $C(r)$ at the
entry of each realization, where it} jumps from the negative plateau of the
disordered side to the positive lobe within a few grid steps. Aligned at the
exit \rev{in Fig.~\ref{fig:supp_boundaries}(b)}, it crosses zero with a much
smaller slope.
\rev{In the odd chains the median slope of $C(r)$ at the entry is
$38$ and $27$, against
$1.2$ and $1.3$ at the exit. The entry
slopes are limited by the field steps, $0.0125$ and
$0.025$, since a jump of $C(r)$ by $0.5$ within one step gives
slopes of $40$ and $20$, so they are lower bounds on the steepness of the
entry, while the exit is resolved by the grid.}
The two crossings also differ in their dependence on the distance at which the
correlator is probed, \rev{shown in Fig.~\ref{fig:supp_boundaries}(c)}. If $C$ is measured at
another odd distance, the entry moves by a median of at most
$0.007$, $0.005$,
$0.013$ and $0.012$, below the grid
steps of $0.0125$ for $L=16,17$ and $0.025$ for $L=32,33$.
The exit moves by up to $0.11$,
$0.06$, $0.20$ and
$0.09$, increasingly as the probing distance moves away
from $L/2$. The entry is a property of the whole chain, while the exit depends
on the scale at which it is measured.

The ground-state energy confirms this difference. By the Hellmann--Feynman
theorem,
\begin{equation}
  -\frac{\partial E}{\partial\Gamma_0}
  =\sum_{i=1}^{L}\frac{\Gamma_i}{\Gamma_0}\langle\sigma^x_i\rangle ,
  \label{eq:supp_hf}
\end{equation}
which we evaluate on the saved states. Figure~\ref{fig:supp_boundaries}(d)
shows the change of this derivative across the crossing, in units of its
typical change per grid step in the same realization. At the entry the median
ratio is $3.1$, $3.2$,
$2.0$ and $1.8$, and
$53\%$, $50\%$,
$11\%$ and $7\%$ of the
entries exceed three. At the exit the median lies between
$0.7$ and $0.9$, and at most
$1\%$ of the exits exceed three. For $\Gamma_i>0$ the
ground state of a finite chain is unique \rev{(Perron--Frobenius)} and its energy is analytic in
$\Gamma_0$, so this is not a true kink. The slope of the energy changes
abruptly at the entry, as at a narrow avoided crossing, while at the exit it
changes at its usual rate.

The exit is a crossing of two magnetizations. With the staggered and uniform
magnetizations per site,
\begin{equation}
  m_s=\frac1L\sum_{i=1}^{L}(-1)^i\langle\sigma^z_i\rangle ,
  \qquad
  m_u=\frac1L\sum_{i=1}^{L}\langle\sigma^z_i\rangle ,
  \label{eq:supp_mag}
\end{equation}
a factorized state with $\langle\sigma^z_i\rangle\simeq m_u+(-1)^i m_s$ gives
$C_{i,i+r}\simeq m_s^2-m_u^2$ for odd $r$ by Eq.~\eqref{eq:cstag}. Within
$0.1$ of the exit the correlation between $C(r)$ and $m_s^2-m_u^2$, over all
realizations and fields, is $0.87$, $0.75$,
$0.85$ and $0.82$, with median deviations
between $0.03$ and $0.07$. Averaged at the exit
\rev{in Fig.~\ref{fig:supp_boundaries}(e)}, $|m_s|$ falls steeply while $m_u$, induced
by the field, grows slowly, and the two curves cross close to the exit for
both parities. Realization by realization, $|m_s|$ falls below $m_u$ after the
entry in $98\%$, $85\%$,
$91\%$ and $81\%$ of the
realizations, at a median distance from the exit of $0.025$,
$0.035$, $0.025$ and
$0.063$, at most three grid steps. Since the
factorization is only approximate, the two criteria agree to this precision
and not better.

\rev{Figure~\ref{fig:supp_boundaries}(f) shows that the exit} is also not a
loss of order throughout the chain. At the exit about half of the pairs at
distance $r$, between $51\%$ and
$54\%$, still have positive correlation, and the strongest
pair has a median correlation between $0.41$ and
$0.61$. The zero of $C(r)$ is the point where the positive and
negative pairs balance on average.

The entry is therefore an abrupt rearrangement of the whole chain, independent
of the probing distance and with an abrupt change in the slope of the energy,
while the exit is a
smooth crossing between the staggered and the uniform magnetization, whose
position depends on the probing distance. Both statements hold for even and
odd chains, so they do not depend on the boundary wall. This is why the
velocities of Sec.~\ref{sec:reentrant} locate the first boundary and not the
second.

\begin{figure*}[t]
  \centering
  \includegraphics[width=\textwidth]{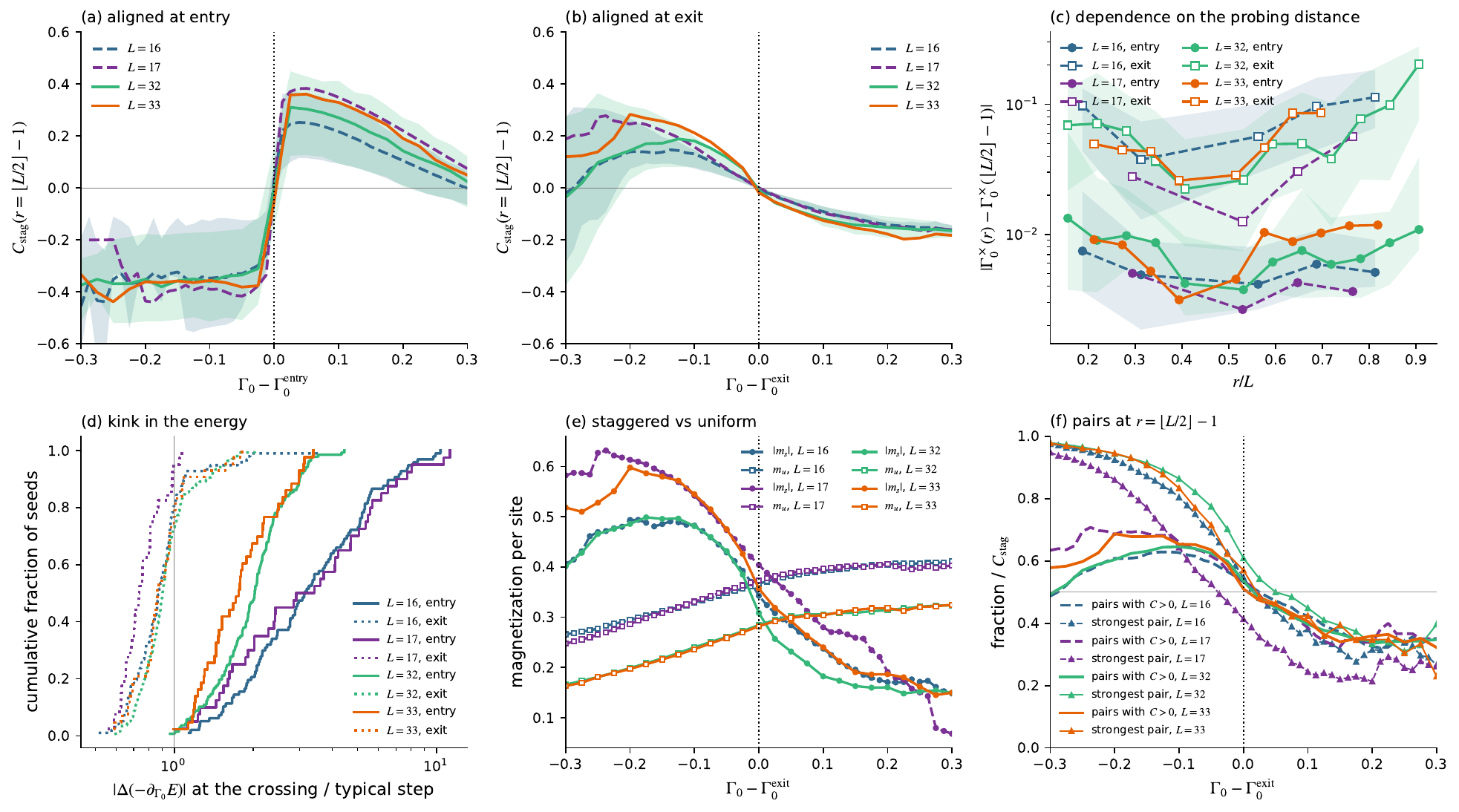}
  \caption{The two boundaries of the reentrant lobe of individual
  realizations, with open boundaries. Even chains $L=16$ ($h=2.5$,
  $97$ reentrant realizations) and $L=32$ ($h=2.3125$,
  $137$), and odd chains without boundary wall, $L=17$ and $L=33$ at
  the same fields ($40$ and $43$). (a),~(b)~Staggered
  correlator at $r=\lfloor L/2\rfloor-1$ aligned at the entry and at the exit
  of each realization (mean, band interquartile range for the even chains).
  (c)~Shift of the entry (filled) and exit (open) \rev{crossing points
  $\Gamma_0^\times$, the zeros of $C(r)$,} when the correlator is
  measured at another odd distance $r$ (median, band interquartile range for
  the even chains). (d)~Cumulative distribution of the change of
  $-\partial E/\partial\Gamma_0$, Eq.~\eqref{eq:supp_hf}, across the entry
  (solid) and the exit (dotted), in units of the typical change per grid step
  of the same realization. (e)~Staggered $|m_s|$ and uniform $m_u$
  magnetization, Eq.~\eqref{eq:supp_mag}, aligned at the exit (mean).
  (f)~Fraction of the pairs at distance $r$ with positive correlation (mean)
  and correlation of the strongest pair (median), aligned at the exit.}
  \label{fig:supp_boundaries}
\end{figure*}

\rev{\smsubsection{Domain walls across the entry}}

The magnetization profiles measured on the saved states identify the domain
walls of each realization directly. A wall is a pair of nearest neighbours
aligned along $z$, both with $|\langle\sigma^z_i\rangle|\ge0.2$, and a site
reverses at the entry when the sign of $(-1)^i\langle\sigma^z_i\rangle$ changes
between the two field values that bracket it. Figure~\ref{fig:supp_walls} shows
one representative realization of each kind of entry. In
$16$ of the $17$ entries in which the total magnetization drops by
two, over the four sizes, a domain of odd length reverses and the two walls
that bound it annihilate, \rev{as in Fig.~\ref{fig:supp_walls}(a)}. Of the $182$
entries that leave the magnetization unchanged, $128$
($70\%$) keep the number of walls, and a block of even length next
to a wall reverses, which displaces the wall, \rev{as in Fig.~\ref{fig:supp_walls}(b)}.
In $23$ of them ($13\%$) two walls that bound a domain of
even length annihilate, \rev{as in Fig.~\ref{fig:supp_walls}(c)}. With the thresholds
$|\langle\sigma^z_i\rangle|\ge0.1$ and $0.3$ the displacements remain the
majority of the entries that keep the magnetization. In the shortest chains a wall can also spread over a few sites
at the entry instead of jumping, which reduces the local magnetization next to
it without reversing any site.
\rev{Table~\ref{tab:supp_walls} gives the complete accounting of the
$317$ entries.}

% Counts from results/review_v1/reentrant-domain-walls-v1/summary.json (events), bins of Delta M of width 1
\begin{table}[t]
  \caption{\rev{Reentrant entries classified by the change $\Delta M$ of the
  total magnetization $\sum_i\langle\sigma^z_i\rangle$, in bins of width one
  centred on the integers, and, for $\Delta M\approx0$, by the change $\Delta W$
  of the number of walls.}}
  \label{tab:supp_walls}
  \begin{ruledtabular}
  \begin{tabular}{lrrrrr}
   & $L=16$ & $L=17$ & $L=32$ & $L=33$ & all \\
  \hline
  \rev{entries} & \rev{97} & \rev{40} & \rev{137} & \rev{43} & \rev{317} \\
  \rev{$\Delta M\approx-2$} & \rev{5} & \rev{2} & \rev{5} & \rev{5} & \rev{17} \\
  \rev{\quad two walls annihilate} & \rev{5} & \rev{2} & \rev{5} & \rev{4} & \rev{16} \\
  \rev{$\Delta M\approx-1$} & \rev{10} & \rev{8} & \rev{20} & \rev{9} & \rev{47} \\
  \rev{$\Delta M\approx0$} & \rev{70} & \rev{24} & \rev{69} & \rev{19} & \rev{182} \\
  \rev{\quad $\Delta W=0$} & \rev{53} & \rev{14} & \rev{48} & \rev{13} & \rev{128} \\
  \rev{\quad $\Delta W=-2$} & \rev{4} & \rev{3} & \rev{15} & \rev{1} & \rev{23} \\
  \rev{\quad other $\Delta W$} & \rev{13} & \rev{7} & \rev{6} & \rev{5} & \rev{31} \\
  \rev{$\Delta M\approx+1$} & \rev{7} & \rev{6} & \rev{36} & \rev{8} & \rev{57} \\
  \rev{$\Delta M\approx+2$} & \rev{5} & \rev{0} & \rev{7} & \rev{2} & \rev{14} \\
  \end{tabular}
  \end{ruledtabular}
\end{table}

% v3 B10: scripts/benchmark_fidelity_reentrant_review_v3.py, results/review_v3/reentrant-fidelity-v1/summary.json (\Rf* macros)
\rev{\smsubsection{Fidelity susceptibility of the reentrant realizations}

We evaluate the fidelity susceptibility $\chi_F$ of every reentrant
realization from the overlaps of the saved states at consecutive fields, with
no new simulation. At the entry it is between $200$ and
$600$ times its median over the scan. Its maximum, with the frozen
rule of the global $H_0$ velocity, lies $0.011$, $0.029$,
$0.023$ and $0.024$ from the physical entry for
$L=16$, $17$, $32$ and $33$. This is closer than the global $H_0$ velocity,
and for $L\ge17$ somewhat farther than the long-range velocity and the slope
of $C(r)$ (Sec.~\ref{sec:reentrant}). For $L=32$ and $33$ the
median overlap of the two states that bracket the maximum is $0.01$
and $0.02$, so the ground state changes almost completely within
one field step, as expected at an avoided crossing between configurations of
domain walls. For $L=16$ it is $0.80$. At the exit $\chi_F$ does not
peak. Its value there is close to its median, only $4$ of the
$97$ realizations with $L=16$ and $18$ of the
$137$ with $L=32$ have an interior second maximum, and its maximum
after the apex of $C(r)$ misses the exit by $0.12$ to $0.22$, more
than a constant equal to the median exit.}

\rev{\smsubsection{Exact diagonalization of the open reentrant chains}

To test whether DMRG with $\chi=32$ stays in metastable configurations of
domain walls, we diagonalize with the Lanczos method all $97$
reentrant realizations with $L=16$ and all $40$ with $L=17$ of
Fig.~\ref{fig:reentrant}, at every field of their scans ($6713$ points). Each point starts
from an independent random vector, so that no state is carried across the
avoided crossing. The DMRG energies agree with the exact ground-state energies
to within $10^{-12}$, and the correlation matrices to within $2\times10^{-10}$. The zeros
of $C(r)$, and hence the entry and the exit, move by less than $10^{-8}$,
and every realization passes the frozen selection rule again. The only
difference is at $\Gamma_0=0$ in one realization with $L=16$, where
two neighbouring couplings add up to $h$, the classical ground state is
twofold degenerate in one spin, and the exact state is \rev{a superposition} of the two
configurations while DMRG keeps one of them. The median of the smallest gap at
the entry is $0.008$ to $0.009$, so the entry is a sharp but resolved
avoided crossing.}

\rev{\smsubsection{Further estimators of the exit}}

\rev{In Fig.~\ref{fig:reentrant}(a) the long-range velocity peaks before the
exit, the second maximum of the global $H_0$ velocity falls after or inside the
physical band, and the first new $H_1$ cycle after the entry, with the rule and
threshold of Sec.~\ref{sec:clean_frontier}, falls inside or before it
depending on the size. For $L=32$ the long-range velocity misses by
$0.079$, the second maximum of the global $H_0$ velocity by $0.096$
and the first $H_1$ cycle by $0.103$, against
$0.071$ for a constant equal to the median exit, and the
odd chains behave in the same way.}

The first $H_1$ cycle after the entry does not locate the exit for any choice
of the lifetime threshold or of the rule. For thresholds between $0.004$
and $0.08$, taking either the first new cycle or the largest increase of
their number, the best realization-resolved error is $0.090$,
$0.063$, $0.085$ and $0.106$ for
$L=16$, $17$, $32$ and $33$, always above that of a constant equal to the
median exit. The slope of the physical correlator misses the exit by
$0.066$ for $L=32$ and $0.103$ for
$L=33$, and the maximum of the half-chain entanglement entropy after the entry
lies $0.090$ and $0.102$ away from
the zero of $C(r)$.

\begin{figure*}[t]
  \centering
  \includegraphics[width=\textwidth]{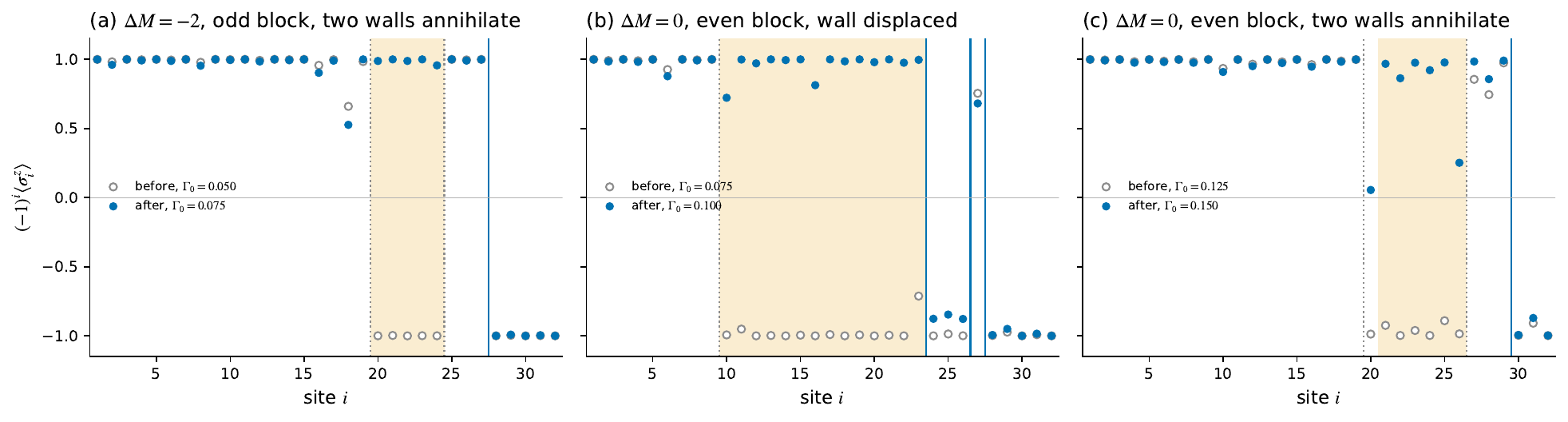}
  \caption{Domain walls across the reentrant entry, for three reentrant
  realizations with $L=32$ at $h=2.3125$. Staggered local magnetization
  $(-1)^i\langle\sigma^z_i\rangle$ at the two field values that bracket the
  entry (open symbols before, filled after). Vertical lines mark the walls
  (dotted before, solid after), and the shaded band the sites that reverse.
  (a)~The magnetization drops by two, a domain of odd length reverses and two
  walls annihilate. (b)~The magnetization is unchanged, a block of even length
  reverses and the wall is displaced. (c)~The magnetization is unchanged, and
  two walls that bound a domain of even length annihilate.}
  \label{fig:supp_walls}
\end{figure*}
\fi

\end{document}